\documentclass{article}

\usepackage{arxiv}

\usepackage[utf8]{inputenc} 
\usepackage[T1]{fontenc}    
\usepackage{hyperref}       
\usepackage{url}            
\usepackage{booktabs}       
\usepackage{amsfonts}       
\usepackage{nicefrac}       
\usepackage{microtype}      
\usepackage{lipsum}		
\usepackage{graphicx}
\usepackage{natbib}
\usepackage{doi}
\usepackage{amsmath}
\usepackage{float}

\def\degree{\hbox{$^\circ$}}

\title{DLESyM-Ocean: A Deep Learning Probabilistic Global Model for Simulating Present-Day Upper Ocean and Sea Ice}

\author{ \href{https://orcid.org/0000-0002-1629-3958}{\includegraphics[scale=0.06]{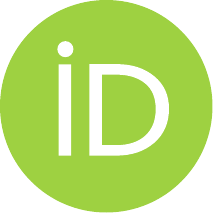}\hspace{1mm}Zachary I. Espinosa} \\
    \thanks{Correspondence to \texttt{zespinosa97@gmail.com}, \texttt{drdee@uw.edu} }
	Department of Atmospheric and Climate Sciences\\
	\And
	\href{https://orcid.org/0000-0002-0600-2870}{\includegraphics[scale=0.06]{orcid.pdf}\hspace{1mm}Nathaniel Cresswell-Clay} \\
	Department of Atmospheric and Climate Sciences\\
	\And
	\href{https://orcid.org/0000-0002-3918-4893}{\includegraphics[scale=0.06]{orcid.pdf}\hspace{1mm}William Yik} \\
	Department of Atmospheric and Climate Sciences\\
	\And
	\href{https://orcid.org/0000-0002-9477-7499}{\includegraphics[scale=0.06]{orcid.pdf}\hspace{1mm}Cecilia M. Bitz} \\
	Department of Atmospheric and Climate Sciences\\
        \And
        \href{https://orcid.org/0000-0002-2608-0868}{\includegraphics[scale=0.06]{orcid.pdf}\hspace{1mm}Edward Blanchard-Wrigglesworth} \\
	Department of Atmospheric and Climate Sciences\\
        \And
        \href{https://orcid.org/0000-0002-4869-6495}{\includegraphics[scale=0.06]{orcid.pdf}\hspace{1mm}Peter Harrington} \\
	NVIDIA \\
        \And
    \href{https://orcid.org/0000-0002-8122-0288}{\includegraphics[scale=0.06]{orcid.pdf}\hspace{1mm}David Pruitt} \\
	NVIDIA \\
        \And
    \href{https://orcid.org/0000-0002-0340-6327}{\includegraphics[scale=0.06]{orcid.pdf}\hspace{1mm}Michael S. Pritchard} \\
	NVIDIA \\
	\And
	\href{https://orcid.org/0000-0002-6390-2584}{\includegraphics[scale=0.06]{orcid.pdf}\hspace{1mm}Dale R. Durran} \\
	Department of Atmospheric and Climate Sciences\\
    NVIDIA \\
}

\renewcommand{\shorttitle}{\textit{arXiv} Template}

\hypersetup{
pdftitle={Espinosa_etal_2026_DLESyM-Ocean},
pdfsubject={physics.ao-ph},
pdfauthor={Zachary Espinosa},
pdfkeywords={Sea Ice, Ocean, Deep Learning, Earth System Modeling},
}

\begin{document}
\maketitle




\begin{abstract}
     While AI has shown remarkable promise in atmospheric and meteorological forecasting, accurately simulating other components of the Earth system with AI remains an active frontier. We present DL{\it ESy}M-Ocean, a Deep Learning Earth System Model that simulates global present-day sea ice and upper ocean conditions. Unlike conventional probabilistic models optimized via diffusion objectives or losses such as continuous-ranked probability score, DL{\it ESy}M-Ocean is trained using a patch energy score loss. When driven by atmospheric forcing, DL{\it ESy}M-Ocean produces a well-calibrated, spatially coherent, and skillful ensemble of sea ice and upper ocean conditions with minimal bias relative to reanalysis products. DL{\it ESy}M-Ocean is stable when autoregressively run for multi-year simulations and produces a climatology and variability with minimal bias compared with reanalysis. We evaluate case studies including a recent sea ice extreme, a severe marine heatwave, the 2023 El Niño transition, and the 2023 spike in global mean temperature. In all of these case studies, DL{\it ESy}M-Ocean produces realistic surface and subsurface trajectories and ample ensemble diversity in response to common atmospheric forcing, suggestive of learned autoregressive ocean dynamics. When coupled with other Earth system components, such as the atmosphere, the computational efficiency of DL{\it ESy}M-Ocean makes it a promising tool for subseasonal to seasonal forecasting.

\end{abstract}



\section{Introduction}
In recent years, artificial intelligence (AI) has emerged as a leading tool for Earth system modeling because it offers a computationally efficient, observation-driven alternative to traditional modeling, with the potential to bypass some structural biases in numerical simulations. To date, atmosphere-only deep learning models have demonstrated skillful probabilistic global weather forecasts out to approximately 15 days \citep{weynCanMachinesLearn2019, keisler2022forecasting, lamGraphCastLearningSkillful2022, kurthFourCastNetAcceleratingGlobal2022, aletSkillfulJointProbabilistic2025}. Remarkably, their performance rivals that of state-of-the-art numerical weather prediction systems and they have begun to be deployed operationally and used in predictability and process understanding studies \citep{vonich2025testing, hakimDynamicalTestsDeep2024}. However, these medium-range weather models typically become unstable in multi-year simulations and lack realistic coupling with evolving boundary conditions from the land surface, ocean, and sea ice \citep{lehmann2026can}. The absence of realistic coupling between components of the Earth system may contribute to the instabilities observed in atmosphere-only deep learning models and poor subseasonal-to-seasonal forecast skill. Previous studies have shown that prescribing sea surface temperatures (SSTs) as lower boundary conditions can enhance the stability of such models \citep{kochkovNeuralGeneralCirculation2024, watt-meyerACE2AccuratelyLearning2024}, enabling decadal-scale rollouts and the emergence of realistic atmospheric trends in response to transient SST forcing\citep{henn2026aimip}. Due to the lack of interactions with other components of the Earth system, atmosphere-only deep learning models are limited to weather prediction and isolated case studies, rendering them unsuitable for investigating internal variability, subseasonal-to-seasonal forecasting, or long-term climate projections.

Recently, ocean-only deep learning models have begun to be developed \citep{subelBuildingOceanClimate2024, arcomanoHybridAtmosphericModel2023, xiongAIGOMSLargeAIDriven2023, bireOceanEmulationFourier2025, holmbergRegionalOceanForecasting2024, holmberg2026njord}. This first generation of models have typically been trained to emulate existing Earth system models (ESMs) and have demonstrated realistic internal variability and long-term stability \citep{dheeshjithSamudraAIGlobal2024}. Similarly, ice-only models have demonstrated state-of-the-art forecast performance in the Arctic at a monthly resolution \citep{anderssonSeasonalArcticSea2021, linIcekNNSouthLightweightMachine2025}.

Hybrid coupled models, combining deep learning and physics-based components, have shown promise in producing stable long-term simulations
\citep{arcomanoHybridAtmosphericModel2023, watt-meyerACE2AccuratelyLearning2024, clarkACE2SOMCouplingSlab2024, kochkovNeuralGeneralCirculation2024, hahner2026representing, yuan2026samudra}, although these models still suffer from some of the limitations and biases of physics-based approaches. Pure deep learning coupled models have also begun to emerge \citep{ wangCoupledOceanAtmosphereDynamics2024,duncanSamudrACEFastAccurate2026}. \cite{cresswell2025deep} recently introduced DL{\it ESy}M, a parsimonious Deep Learning Earth System Model that couples separate atmosphere and ocean components to produce stable, 1000-year simulations of Earth's present-day climate. The model exhibits realistic seasonal and interannual variability, comparable to and in some cases outperforming CMIP6 historical simulations. While DL{\it ESy}M marks a significant step forward in data-driven Earth system modeling, its ocean component remains highly simplified, deterministic and limited to simulating sea surface temperatures. This simplification likely contributes to DL{\it ESy}M’s muted ocean variability, stemming from the absence of upper-ocean heat content and ocean dynamics. Additionally, DL{\it ESy}M does not simulate sea ice, a critical feature of high-latitude, regional weather and global climate, or subsurface ocean conditions.

Here, we build upon the work of \cite{cresswell2025deep} and \cite{karlbauer2024advancing} and introduce DL{\it ESy}M-Ocean, a probabilistic autoregressive Deep Learning Earth System Model for simulating global present-day upper-ocean and sea ice conditions. When driven by atmospheric forcing, DL{\it ESy}M-Ocean produces a well-calibrated and skillful ensemble of sea ice and upper ocean conditions. DL{\it ESy}M-Ocean is trained on ERA5 and UFS-Replay reanalysis and is stable when autoregressively run for multiple years, generating realistic sea ice and upper ocean climatology and variability. Using ensemble predictions, we assess DL{\it ESy}M-Ocean's ability to reproduce recent marine heatwaves, ENSO, and Arctic and Antarctic sea ice extremes. DL{\it ESy}M-Ocean represents an important step towards developing skillful deep learning-based Earth system models that can be used for probabilistic subseasonal-to-seasonal prediction and climate modeling.

\section{Model Design}
\label{sec:design}


\subsection{Architecture}
DL{\it ESy}M-Ocean largely uses the same architecture as the ocean module in \cite{cresswell2025deep}, with a U-Net style convolutional neural network, diagrammed in Figure \ref{fig:figs1}. It uses two sequential ConvNext blocks at each level of the U-Net and three total levels, with channel depths of $D_1 = 136$, $D_2 = 64$, and $D_3 = 34$. Two notable differences from the architecture of \cite{cresswell2025deep} are the network's upsampling and HEALPix padding schemes. We find that the transposed convolutions used for upsampling in \cite{karlbauer2024advancing} and \cite{cresswell2025deep} introduce checkerboarding artifacts during inference, a well documented problem in the computer vision literature \citep{odena2016deconvolution}. Simpler upsampling methods such as nearest neighbor and bilinear interpolation also introduce checkerboarding artifacts when applied na\"ively to HEALPix data. We use a custom upsampling technique which sequentially applies nearest-neighbor upsampling, a four-point smoother, and a small convolutional layer. This configuration mitigates checkerboard artifacts while preserving the learning capacity typical of transposed convolutions. For the HEALPix padding scheme, we find that the padding method introduced by \cite{karlbauer2024advancing} and subsequently used by \cite{cresswell2025deep} and \cite{brenowitz2025climate} introduces artifacts along the seams of the polar faces of the HEALPix grid. This is because pixels along these seams were not padded with pixels from the same latitude ring, unlike pixels in the interior of the faces. We address this by using an ``isolatitude" padding scheme which preserves zonal gradients and eliminates HEALPix face seam artifacts. In total, DL{\it ESy}M-Ocean is quite small relative to other AI Earth System models, with only ~3.1M trainable parameters.

DL{\it ESy}M-Ocean ingests two time frames comprising current ($x_0$) and previous initial conditions ($x_{-96}$) to predict sequential ocean and sea ice states at 4-day ($x_{96}$) and 8-day ($x_{192}$) lead times. The model uses a global skip connection to the most recent observed state ($x_0$), effectively formulating the objective as  learning the residual $(\delta x_{96}, \delta x_{192})$ such that $(x_{96},x_{192})=(x_{-96},x_{0})+(\delta x_{96}, \delta x_{192})$.

Residual prediction leverages the high temporal autocorrelation inherent in ocean and ice variables. To prevent physically impossible states such as negative ice thickness or sea ice concentrations exceeding 100\%, we enforce a hard boundary constraint during the forward pass immediately after the residual connection with $x_0$. The boundary constraint clips the normalized predictions to guarantee that the sea ice concentration (sic) and sea ice thickness (sit) model outputs are in valid physical ranges (0 $\leq$ sic $\leq$ 1 and sit $\geq$ 0). This is formulated as: 
\begin{equation}
   \hat{x}'_v = \max\left(\frac{b_{\min} - \mu_v}{\sigma_v}, \min\left(\hat{x}_v, \frac{b_{\max} - \mu_v}{\sigma_v}\right)\right)
\end{equation}
 where $\hat{x}_v$ is the normalized network prediction for a variable $v$ with training mean $\mu_v$, standard deviation $\sigma_v$, and physical limits $[b_{\min}, b_{\max}]$. 
 
In order to produce ensemble forecasts, we inject a global noise vector $n_v \sim N(0,1)^{32}$ into the conditional layer norms (CLNs) at the beginning of each ConvNeXt block in the last levels of the U-Net's encoder and decoder (Figure \ref{fig:figs1}). The CLN uses multi-layer perceptrons (MLPs) to project the conditioning noise vector into channel-wise scale ($\gamma$) and shift ($\beta$) parameters. Each member in the ensemble forecast is  generated by sampling a different global noise vector. Given an input tensor $X \in \mathbb{R}^{B \times C \times H \times W}$ and a conditioning noise vector $n_v$, the CLN operator applies a channel-wise normalization followed by an affine transformation parameterized by $n_c$:

\begin{equation}
Y = \left( c + \text{MLP}_\gamma(Z) \right) \odot \text{LayerNorm}(X) + \text{MLP}_\beta(Z)
\end{equation}

where $\text{LayerNorm}(X) = \frac{X - \mu}{\sqrt{\sigma^2 + \epsilon}}$ computes the standard normalization over the channel dimension. A small constant, $\epsilon=1e-5$, is added to prevent division by zero; both MLPs use 2 64-dimensional layers, and $c=1.0$, is a constant scaling center. The weights of the last MLP layers are initialized to 0, simulating a deterministic model in early training. We find this empirically to improve training stability and accelerate convergence. 

\subsection{Almost Fair Patch Energy Score Loss Function}
Probabilistic models generate forecast distributions rather than deterministic point predictions, necessitating strictly proper scoring rules to penalize the statistical divergence between the predicted distribution and the observed target. The Continuous Ranked Probability Score (CRPS), for empirical distributions \cite{hersbach2000decomposition}, is defined as:

\begin{equation}
    CRPS(\{x_i\}_{i=1}^M,y) = \frac{1}{M} \sum_{i=1}^M |x_i - y| - \frac{1}{2M^2}\sum_i^M\sum_j^M|x_i - x_j|
\end{equation} 

where $y$ is the target and $x_i$ is the $i^{th}$ prediction from forecast ensemble of size $M$. CRPS penalizes both ensemble spread and deviations of the ensemble mean from the verification. Because CRPS evaluates marginal distributions, it does not enforce spatial coherence within individual ensemble members, often resulting in spatial artifacts. Atmospheric models frequently mitigate this by supplementing CRPS with a spectral Fourier loss \citep{bonev2025fourcastnet}; however, complex continental boundaries make the application of Fourier spectral methods to ocean domains challenging. 

Here, we use Energy Score (ES), which generalizes CRPS to evaluate multivariate joint distributions and improves spatial coherence. For an empirical ensemble, forecasting a $d$-dimensional vector target $\mathbf{y} \in \mathbb{R}^d$, it is defined as:

$$ES(\{\mathbf{x}_i\}_{i=1}^M,\mathbf{y}) = \frac{1}{M} \sum_{i=1}^M \|\mathbf{x}_i - \mathbf{y}\|_2 - \frac{1}{2M^2}\sum_{i=1}^M\sum_{j=1}^M\|\mathbf{x}_i - \mathbf{x}_j\|_2$$

where $\|\cdot\|_2$ denotes the Euclidean ($L_2$) norm. For scalar variables ($d=1$), the ES reduces exactly to CRPS. Because distance concentration in high-dimensional spaces renders the global Euclidean norm insensitive to localized errors \citep{beyer1999nearest, aggarwal2001surprising}, computing the ES over highly multivariate global domains causes training instability and poor convergence. To enforce spatial coherence while maintaining stable convergence, we restrict the ES evaluation to localized spatial patches, defining the Patch Energy Score (PES):

$$PES(\{\mathbf{x}_{i,p}\}_{i=1}^M,\mathbf{y}_p) = \frac{1}{M} \sum_{i=1}^M \|\mathbf{x}_{i,p} - \mathbf{y}_p\|_2 - \frac{1}{2M^2}\sum_{i=1}^M\sum_{j=1}^M\|\mathbf{x}_{i,p} - \mathbf{x}_{j,p}\|_2$$

where $p$ indexes a $3 \times 3$ spatial patch. To avoid cross-face padding, patches are confined to individual HEALPix faces. We apply a global stride of 1 to create dense, overlapping coverage.

Although the PES is a strictly proper scoring rule, the standard empirical formulation is a biased estimator for finite ensemble sizes. The strictly fair estimator \citep{ferro2014fair, leutbecher2017stochastic} corrects this finite-sample bias:

$$fPES(\{\mathbf{x}_{i,p}\}_{i=1}^M,\mathbf{y}_p) = \frac{1}{M} \sum_{i=1}^M \|\mathbf{x}_{i,p} - \mathbf{y}_p\|_2 - \frac{1}{2M(M-1)}\sum_{i=1}^M\sum_{j=1}^M\|\mathbf{x}_{i,p} - \mathbf{x}_{j,p}\|_2$$

However, the fair estimator is characterized by several key degeneracies which degrade training dynamics (refer to Supplemental text for an in-depth discussion). To resolve these degeneracies, we adopt an "almost-fair" parameterization, similar to almost-fair CRPS \citep{lang2026aifs}, which interpolates between the biased and fair estimators using a tuning parameter $\alpha \in [0, 1]$:

$$afPES(\{\mathbf{x}_{i,p}\}_{i=1}^M,\mathbf{y}_p) = \frac{1}{M} \sum_{i=1}^M \|\mathbf{x}_{i,p} - \mathbf{y}_p\|_2 - \frac{1 - \epsilon}{2M(M-1)}\sum_{i=1}^M\sum_{j=1}^M\|\mathbf{x}_{i,p} - \mathbf{x}_{j,p}\|_2$$

where $\epsilon = \frac{1 - \alpha}{M}$.

Finally, to apply the almost-fair PES across patches containing land and ocean pixels, we incorporate a binary land-sea mask alongside a Horvitz-Thompson-style extrapolation. For a spatial patch $p$ containing $P^2$ total cells and $n_{valid,p}$ active ocean cells, the masked, extrapolated Euclidean distance is computed as:

$$\|\mathbf{x}_{i,p} - \mathbf{y}_p\|_{extrapolated} = \sqrt{ \frac{P^2}{\max(1, n_{valid,p})} \sum_{k=1}^{P^2} v_{p,k} (x_{i,p,k} - y_{p,k})^2 }$$

where $v_{p,k} \in \{0, 1\}$ represents the land-sea mask at cell $k$. This formulation ensures that the $L_2$ norm magnitude scales consistently regardless of land-sea geometry. Fully masked terrestrial patches ($n_{valid,p} = 0$) compute to zero and are excluded from the spatial reduction.

\subsection{Training Procedure}
We train DL{\it ESy}M-Ocean using the following two stages: 

Stage 1: 1500 epochs using 3 AR steps and $\alpha=0.95$ (afPES) \\
Stage 2: 600 epochs using 4-10 AR steps (100 epoch per AR step) and $\alpha=0.95$ (afPES) \\

 We use an ensemble size of $M=2$. The total afPES loss is then the average of afPES globally and over all timesteps. In stage 1, we perform three autoregressive (AR) forward passes through DL{\it ESy}M-Ocean to generate a 24-day forecast. The loss is computed on the final output step (days 20 and 24), and backpropagation is applied through all three AR steps. During stage 2, we progressively increase the number of AR steps, and find that this finetuning substantially improves model stability and forecast skill at seasonal lead times. We use gradient checkpointing to accommodate this long horizon training. Refer to Table \ref{tab:tableS2} in the Supporting Information for additional details on each training stage. 

As in \cite{karlbauer2024advancing} and \cite{cresswell2025deep} we scale the weight for each variable (e.g. $w_i$) using $w_i = 0.001/V_i$ where $V_i$ is the unscaled validation loss after a 10-epoch training. This per-variable weighting ensures that each variable contributes roughly equally to the loss. We use adaptive gradient clipping by independently bounding each output unit's gradient $L_2$ norm relative to its corresponding weight norm, and we skip low-dimensional parameters like biases. For each unit $i$, the gradient $G_i$ is scaled as \citep{brock2021high}: 

\begin{equation}
 G'_i = G_i \min\left(1, \frac{\lambda \max(||W_i||_2, \epsilon)}{||G_i||_2}\right)   
\end{equation} where $\lambda=0.05$ is the clip\_factor and $\epsilon=0.001$ is a stabilizing floor. During stage 2, following an initial warmup epoch ($t_{start}=1$), we maintain a copy of the model parameters updated via an exponential moving average (EMA): 
\begin{equation}
    \theta_{\text{EMA}}^{(t)} = \alpha \theta_{\text{EMA}}^{(t-1)} + (1 - \alpha) \theta^{(t)}
\end{equation}
where $\theta^{(t)}$ represents the current online model weights and $\alpha=0.999$ is a decay factor. We empirically find EMA reduces checkpoint variability and initial condition sensitivity \citep{polyak1992acceleration, tarvainen2017mean}, and use the final EMA weights for our model simulations. 

\subsection{Data}
DL{\it ESy}M-Ocean is trained on ERA5 reanalysis \citep{hersbach2020era5} and UFS-Replay \citep{orbe2017large, gichamo2022optimal} data from 1994–2018, validated using data from 2019–2021, and tested on data from 2022–2023. The UFS-Replay dataset is a high-resolution, global coupled reanalysis-style product generated by `replaying' the Unified Forecast System (UFS) coupled model, comprising atmosphere (FV3), ocean (MOM6), sea ice (CICE6), and land (Noah-MP) components, to the ERA5 and ORAS5 reanalysis datasets \citep{orbe2017large}. The DL{\it ESy}M-Ocean training dataset consists of four types of inputs: 23 prognostic fields, 2 constant fields (topographic height and land-sea fraction), 1 prescribed field (top of atmosphere incoming solar radiation), and 3 coupled atmospheric fields (10-m windspeed, 1000-hPa height and outgoing longwave radiation). Surface air temperature, which is highly correlated with SST, is not input as a coupled field to prevent the model from learning relationships that lack physical causality. Refer to Table \ref{tab:tableS1} in the Supporting Information for a full list of input and output variables for DL{\it ESy}M-Ocean.

Consistent with \cite{cresswell2025deep}, all raw 0.25$\degree$ data are regridded from their native coordinate systems to the Hierarchical Equal Area isoLatitude Pixelization (HEALPix; \citep{gorskiHEALPixFrameworkHighResolution2005, zonca2019healpy, karlbauer2024advancing}). The HEALPix mesh uses twelve equal-area base faces, with four lying in the tropics and four in each pole. Each base face is further subdivided into 64x64 cells, resulting in 49,152 cells globally. This corresponds to a nominal resolution of approximately 1$\degree$ at the equator, with an average distance of $\sim$110 km between adjacent cell centers. Consistent with \cite{cresswell2025deep} all ocean variables are imputed using zonal linear interpolation across land, in order to ensure that spatial convolutions generate reasonable values near land-sea boundaries. During inference, the global ocean variables are allowed to freely evolve. Immediately after inference and prior to evaluation, all data are regridded to a common 1$\degree$ global grid, and a binary land-sea mask is applied to all ocean variables using an ocean fraction $\geq 0.50$.

\section{Results}
\subsection{Driven Forecast Skill}
\label{sec:forecast-skill}


Using DL{\it ESy}M-Ocean, we generate 90-day, 50-member ensemble ``forecasts" initialized weekly from January 1, 2022, through December 31, 2022. To evaluate the influence of atmospheric conditions, the model is driven by two distinct forcings. The first configuration drives DL{\it ESy}M-Ocean with ERA5 atmospheric reanalysis corresponding to the forecast valid time. 
This configuration allows us to isolate the performance of the ocean model without the compounding errors that would be introduced  from free running atmospheric forecasts, but 
does not reflect DL{\it ESy}M-Ocean's operational predictive skill because it uses perfect atmospheric forcing well beyond the limit of intrinsic atmospheric predictability. Unless otherwise stated, all DL{\it ESy}M-Ocean simulations in subsequent analysis use this configuration of atmospheric forcing. The second setup drives the model with an ERA5 daily atmospheric climatology computed for each calendar date.  As the model is stepped forward, it is forced by the climatological atmospheric conditions for each successive calendar day; we refer to results obtained using this atmospheric climatology as DL{\it ESy}M-Ocean-AC. We benchmark model performance against two standard reference forecasts: a persistence baseline, which projects the initial ocean state uniformly across all future lead times without evolution, and a 25-member probabilistic ocean climatology, formulated by sampling the corresponding historical ocean states from each individual year between 1994 and 2019 as discrete ensemble members. Evaluation metrics are computed using WeatherBenchX \citep{raspWeatherBenchBenchmarkNext2023} and unbiased estimators are chosen when available.

\begin{figure}\centering\noindent\includegraphics[width=\textwidth,angle=0]{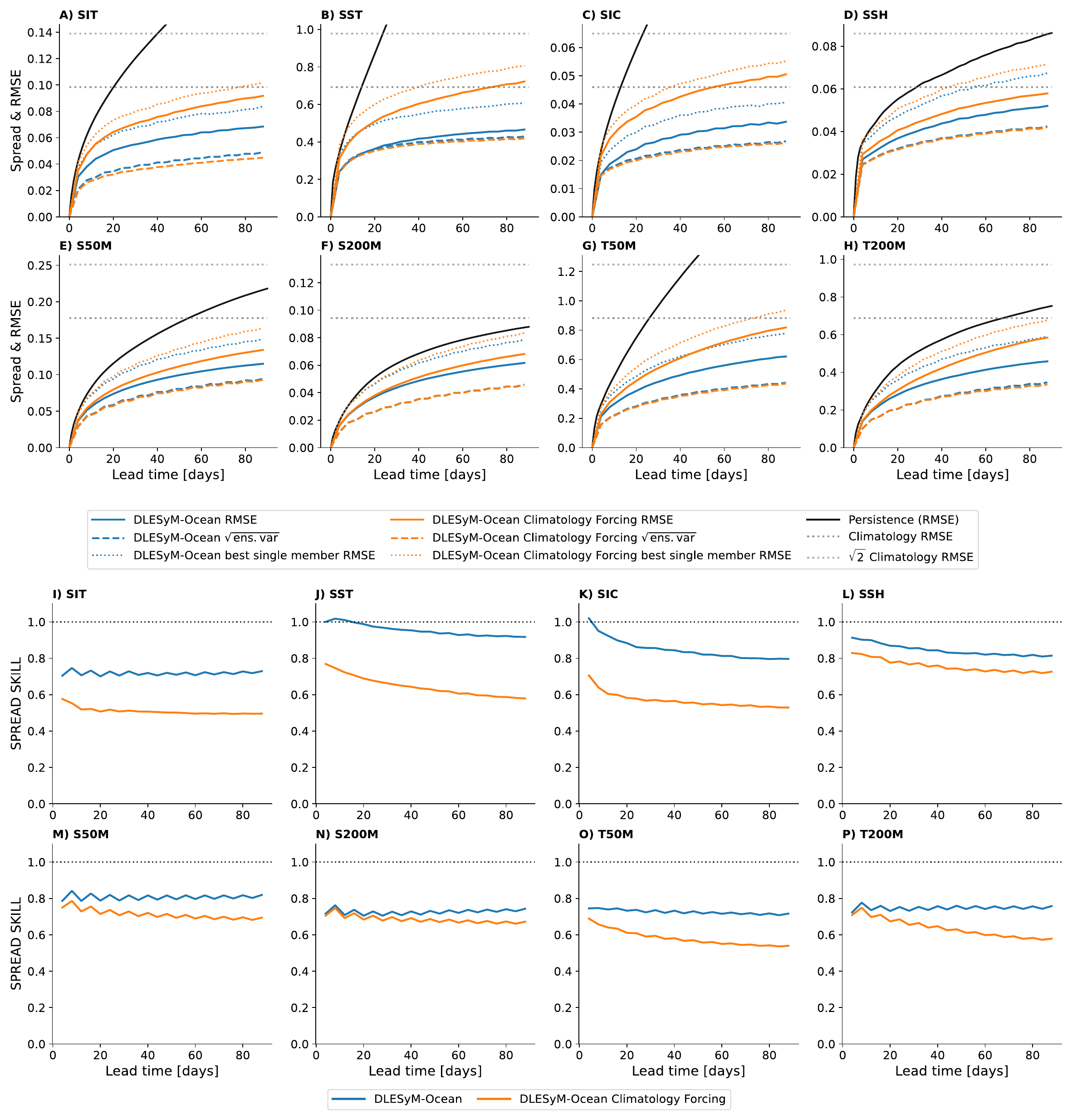}\\
 \caption{\textbf{Forecast Skill:} (A–H) Ensemble spread (dashed lines) and root-mean-square error (RMSE) of the ensemble mean (solid lines) and a single ensemble member (dotted lines) for 50-member, 90-day DL{\it ESy}M-Ocean forecasts initialized weekly from 2022-01 through 2022-12. Forecasts are shown for two configurations: DLESyM-Ocean driven by concurrent ERA5 atmospheric forcing (blue) and DLESyM-Ocean-AC (orange). Comparison benchmarks include RMSE for persistence (solid black line) and the mean of a 25-member probabilistic climatology drawn from 1994–2019 (dotted gray line).  Top gray dotted line is $\sqrt{2}$ times this climatological RMSE. Panels display (A) sea ice thickness (SIT), (B) sea surface temperature (SST), (C) sea ice concentration (SIC), (D) sea surface height (SSH), (E) salinity at 50 m depth (S50M), (F) salinity at 200 m depth (S200M), (G) temperature at 50 m depth (T50M), and (H) temperature at 200 m depth (T200M). (I–P) Unbiased spread-skill ratio (SSR) as a function of lead time over the 90-day forecast period for both configurations (blue and orange lines as in A–H). Panels (I–P) correspond to the same variables listed in (A–H), respectively.}
 
 \label{fig:fig1}
\end{figure}

We find that DL{\it ESy}M-Ocean exhibits substantially lower ensemble mean root-mean-square error (RMSE\_ENS) than both persistence and ocean climatological forecasts for all variables and at nearly all lead times under both atmospheric forcing configurations (Figure \ref{fig:fig1}A-H; Figure \ref{fig:figs3} in SI). As expected, driving DLESyM-Ocean with concurrent ERA5 atmospheric fields yields greater forecast skill than DLESyM-Ocean-AC. The performance difference between these configurations is most pronounced for surface variables, relative to subsurface variables, reflecting fundamental differences in their timescale of evolution and the role of atmospheric forcing. 

RMSE\_ENS should asymptote to the RMSE error of the climatological mean forecast, while RMSE errors for individual forecasts should asymptote to a value $\sqrt{2}$ times that mean.  As apparent in Fig.~\ref{fig:fig1}A-H, none of the RMSE errors in rollouts forced by concurrent ERA5 fields grow to their theoretical asymptotic values over the 90-day rollout driven by the correct ERA5 atmospheric fields. This excellent performance would undoubtedly be degraded in a coupled simulation in which atmospheric fields were also forecast and the error in those atmospheric fields saturated as expected after about two weeks. A rough measure of the model performance that might be expected once the errors in the atmospheric fields saturate is provided by the DLESyM-Ocean-AC;  even when climatology replaces the correct atmospheric conditions, the 90-day RMSE\_ENS remains below that for the RMSE of an ocean climatological estimate of all the target variables except SST and SIC. Neither the RMSE\_ENS of the SST and SIC (which are too high), nor their RMSE from single  ensemble member rollouts (which are too low) are asymptoting to the correct values at the end of the 90-day rollouts.  In those cases where the RMSE of the ensemble mean fields at deeper levels remain well below climatology, such as S200m and T200m, the simulation may simply not have had enough time for the values to grow to their asymptotic limit.  

In a well calibrated ensemble, the joint distribution of the ensemble members is unchanged if one member is exchanged with the verifying observation. When this condition holds, the Spread-Skill Ratio (SSR),  
defined as the ratio of ensemble spread (calculated per-variable as the spatially averaged square root of the MSE across ensemble members) to the RMSE\_ENS is unity \citep{fortin2014should}. SSR values greater (less) than 1 denote an overdispersive (underdispersive) system wherein the spread is excessively large (small) relative to the forecast error magnitude. When driven by the correct atmospheric forcing, DL{\it ESy}M-Ocean is modestly underdispersive for most variables and lead times (Figures~\ref{fig:fig1}I-P, \ref{fig:figs5}), as is also evident from the rank histograms plotted in Figure~\ref{fig:figs6}. When driven by the climatological atmospheric variables the degree of underdispersion is worse, which might be expected since these climatological fields are smoother than those on any given day, and it would not be possible to exchange one of them with the observed atmospheric fields without changing the joint distribution of the ensemble forcing.
 
\begin{figure}\noindent\includegraphics[width=\textwidth,angle=0]{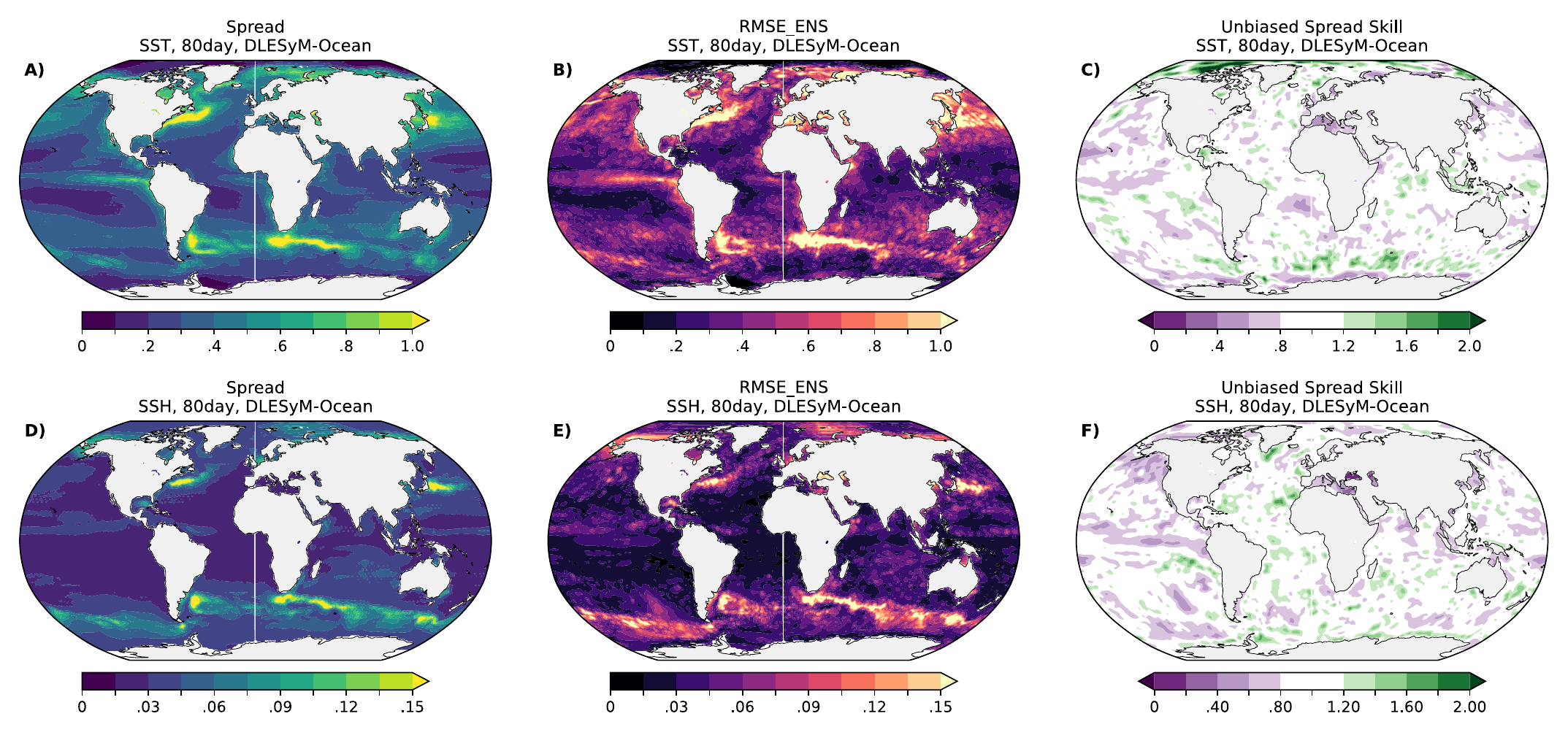}\\
 \caption{\textbf{Spread \& Calibration:} (A–C) Spatial distributions of (A) ensemble spread, (B) ensemble-mean RMSE (RMSE\_ENS), and (C) the unbiased spread-skill ratio for SST, averaged across all initialization times at an 80-day lead time. (D–F) As in (A–C), but for SSH. As in Figure 1, we account for the finite ensemble size ($N=50$) by calculating the ensemble variance with Bessel's correction ($N-1$) and removing the positive bias in the ensemble-mean MSE by subtracting the internal ensemble variance scaled by $1/N$.}\label{fig:fig2}
\end{figure}

Some degree of underdispersion is not strictly a deficiency for a probabilistic model like DL{\it ESy}M-Ocean that generates ensemble variability entirely from internal noise designed to capture the aleatoric uncertainty stemming from inherently chaotic oceanic dynamics. The forecasts evaluated here do not attempt to represent epistemic uncertainty, due to imperfect initial conditions and model bias. Incorporating initial condition perturbations, a well-established methodology outside the immediate scope of this study \citep{hamillComparisonProbabilisticForecasts2000, tangENSOPredictabilityFully2006}, would likely increase total ensemble spread and could improve calibration. To investigate DL{\it ESy}M-Ocean spread in greater detail, we compute time-averaged fields of ensemble spread, RMSE\_ENS, and SSR at an 80-day lead time across all initializations (Figure \ref{fig:fig2}). Although DL{\it ESy}M-Ocean is generally underdispersive, the spatial distribution of ensemble spread strongly correlates with RMSE\_ENS ($r = 0.85$), indicating that DL{\it ESy}M-Ocean scales uncertainty in regions with lower intrinsic predictability. Maximum spread and error occur within regions dominated by mesoscale and submesoscale eddy activity, such as the Gulf Stream, Kuroshio Current, and Antarctic Circumpolar Current.

\subsection{Climatology: Ocean and Sea Ice}
\label{sec:ocean-ice-clim}

\begin{figure}\noindent\includegraphics[width=.9\textwidth,angle=0]{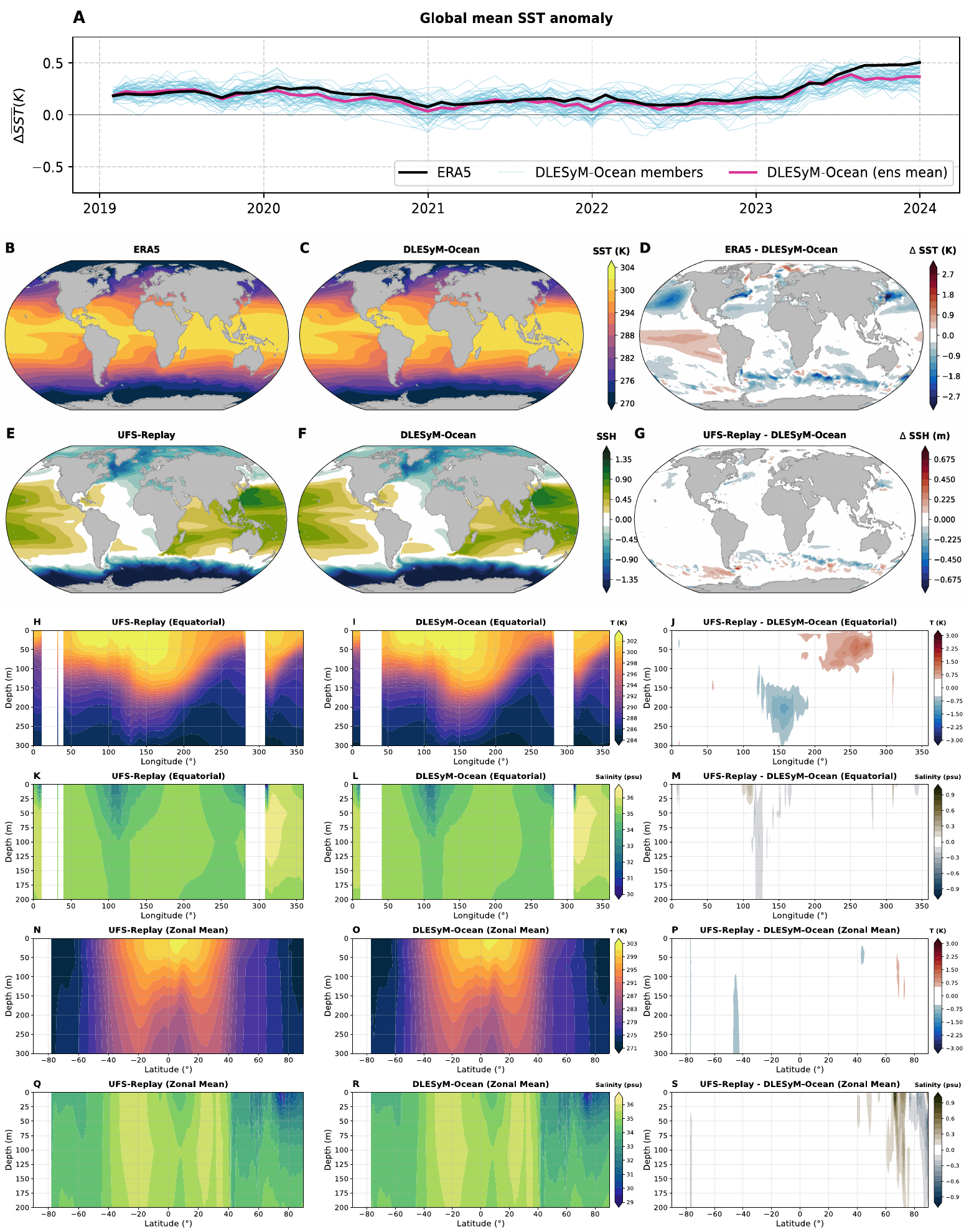}\\
\centering
 \caption{\textbf{Global Climatology:}  (A) Time series of the global mean sea surface temperature (SST) anomaly for ERA5 (solid black line), DL{\it ESy}M-Ocean ensemble mean (solid magenta line), and DL{\it ESy}M-Ocean ensemble members (solid blue lines; $N=50$) from January 2019 through December 2023, computed relative to a 1994–2018 monthly ERA5 climatology. (B–D) Global 5-year mean spatial distributions of SST over the 2019–2023 period for (B) ERA5, (C) DL{\it ESy}M-Ocean, and (D) their difference (ERA5 minus DL{\it ESy}M-Ocean). (E–G) As in (B–D), but for sea surface height (SSH). (H–J) Equatorial (5$\degree$S–5$\degree$N) meridionally averaged depth-longitude cross sections of subsurface temperature for (H) UFS-Replay, (I) DL{\it ESy}M-Ocean, and (J) their difference (UFS-Replay minus DL{\it ESy}M-Ocean). (K–M) As in (H–J), but for salinity. (N–P) Global zonal mean (90$\degree$S–90$\degree$N) depth-latitude cross sections of subsurface temperature for (N) UFS-Replay, (O) DL{\it ESy}M-Ocean, and (P) their difference. (Q–S) As in (N–P), but for salinity.}\label{fig:fig3}
\end{figure}

To evaluate multi-year stability, we generate a 5-year, 50-member integration of DL{\it ESy}M-Ocean spanning the validation and test periods. The model is initialized with the ocean state on January 1, 2019 and driven by atmospheric conditions through December 2023, allowing the ocean variables to freely evolve over 228 autoregressive steps. We find that the DL{\it ESy}M-Ocean ensemble mean maintains a realistic global climatology for both SST and SSH with minimal drift (Figure \ref{fig:fig3}B-G). In the equatorial Pacific, the model reproduces the mean state zonal gradients and characteristic subsurface temperature and salinity structure. DL{\it ESy}M-Ocean captures the shoaled thermocline in the eastern Pacific cold tongue, the deep Western Pacific Warm Pool (Figure \ref{fig:fig3}H–J), and the western Pacific fresh pool (Figure \ref{fig:fig3}K–M). The model generates realistic meridional pole-to-equator temperature and salinity gradients, exhibiting only a minor subsurface salinity bias in the northern high latitudes (Figure \ref{fig:fig3}N–S). The most pronounced biases are localized within coastal regimes, likely resulting from unresolved subgrid-scale boundary processes and artificial land-sea masking effects, and in regimes characterized by strong eddy kinetic energy and low predictability, such as the Southern Ocean and Gulf Stream. Over the 5-year period, the model reproduces global interannual SST variability, with the reanalysis anomalies falling within the DL{\it ESy}M-Ocean 50-member ensemble spread (Figure \ref{fig:fig3}A). DL{\it ESy}M-Ocean captures the rise in global mean temperatures observed during 2023 though the ensemble mean slightly underestimates the late-2023 temperature peak \citep{blanchard2025record}. Given that DL{\it ESy}M-Ocean is forced by atmospheric OLR, the bias in peak global mean temperature likely stems from a global energy bias caused by internal model drift.

\begin{figure}\noindent\centering\includegraphics[width=.95\textwidth,angle=0]{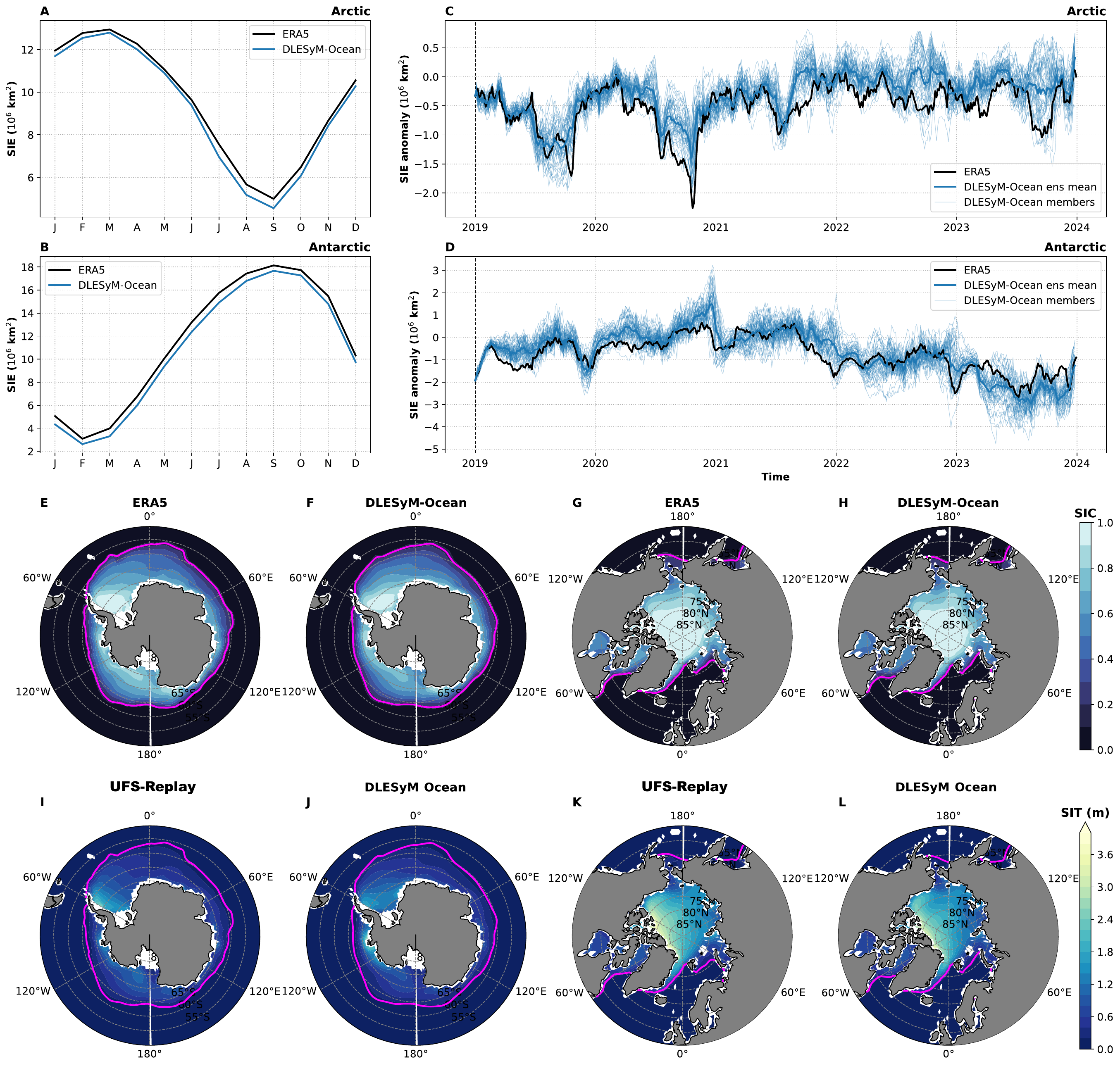}\\
 \caption{\textbf{Sea Ice Mean Climatology and Monthly Variability:} Annual cycle of Arctic (A) 
 and Antarctic (B) 
 sea ice extent (SIE) averaged over 2019-01 to 2023-12 for ERA5 and DL{\it ESy}M-Ocean. (C) Time series of Arctic SIE anomaly for ERA5 (solid black line), DL{\it ESy}M-Ocean ensemble mean (solid magenta line), and DL{\it ESy}M-Ocean ensemble members (thin solid blue lines; $N=50$) between 2019-01-01 and 2023-12-31 for (C) the Arctic and (D) the Antarctic. Spatial distributions of 5-year annual mean SIC for ERA5 and DL{\it ESy}M-Ocean ensemble mean in the Antarctic (E,F) and the Arctic (G,H). Magenta contours indicate the sea ice edge (SIC = 0.15). As in (E–H), but for sea ice thickness (I-L).}
 \label{fig:fig4}
\end{figure}

Next, we examine Arctic and Antarctic sea ice in DL{\it ESy}M-Ocean. When driven by concurrent ERA5 atmospheric variables from 2019 through 2023, the 5-year-average ensemble-mean annual cycles of sea ice extent (SIE, defined as the total area of all grid cells with SIC $\geq$ 15\%) exhibit a very small positive bias in the Arctic and virtually no error in the Antarctic (Figure \ref{fig:fig4}A,B). Moreover, the 5-year annual average ensemble-mean spatial distributions of sea ice concentration (SIC) and sea ice thickness (SIT) in DL{\it ESy}M-Ocean closely resemble those in ERA5 (Figure~\ref{fig:fig4}E-L). 
To evaluate monthly variability, we calculate SIE anomalies between January 2019 and December 2023 using the 5-year autoregressive rollout of the  50-member  DL{\it ESy}M-Ocean ensemble (Figure \ref{fig:fig4}C-D). The anomalies are calculated using the ERA5 4-day SIE climatology from 1994 and 2018. The model reproduces monthly SIE variability, with some ensemble members capturing both the late-2020 Arctic and late-2022 Antarctic SIE extremes. The enhanced ensemble spread of SIE during summer and fall in both hemispheres indicates that DL{\it ESy}M-Ocean correctly captures internal variability in the intrinsic ocean and sea ice dynamics under common atmospheric forcing. While the spread in the 50-member ensemble generally encompasses the reanalysis anomalies, the ensemble mean diverges from reanalysis by roughly 0.2 K for the last six months, making it unclear whether the reanalysis represents an extreme realization of internal variability or if the ensemble develops a systematic bias.

\subsection{Variability: El Niño}
\label{sec:enso}

The El Niño-Southern Oscillation (ENSO) is the leading mode of upper ocean variability on interannual time scales. To assess DL{\it ESy}M-Ocean's ability to reproduce historical ENSO patterns over a long period of record, we generate an autoregressive 16-member ensemble beginning on January 1, 1994 and continuing through 2022. To utilize a long record of atmospheric forcing, this period does include the training data, but the longest autoregressive training periods are 80 days, far shorter than these 29-year rollouts.  

\begin{figure}\centering\noindent\includegraphics[width=.9\textwidth,angle=0]{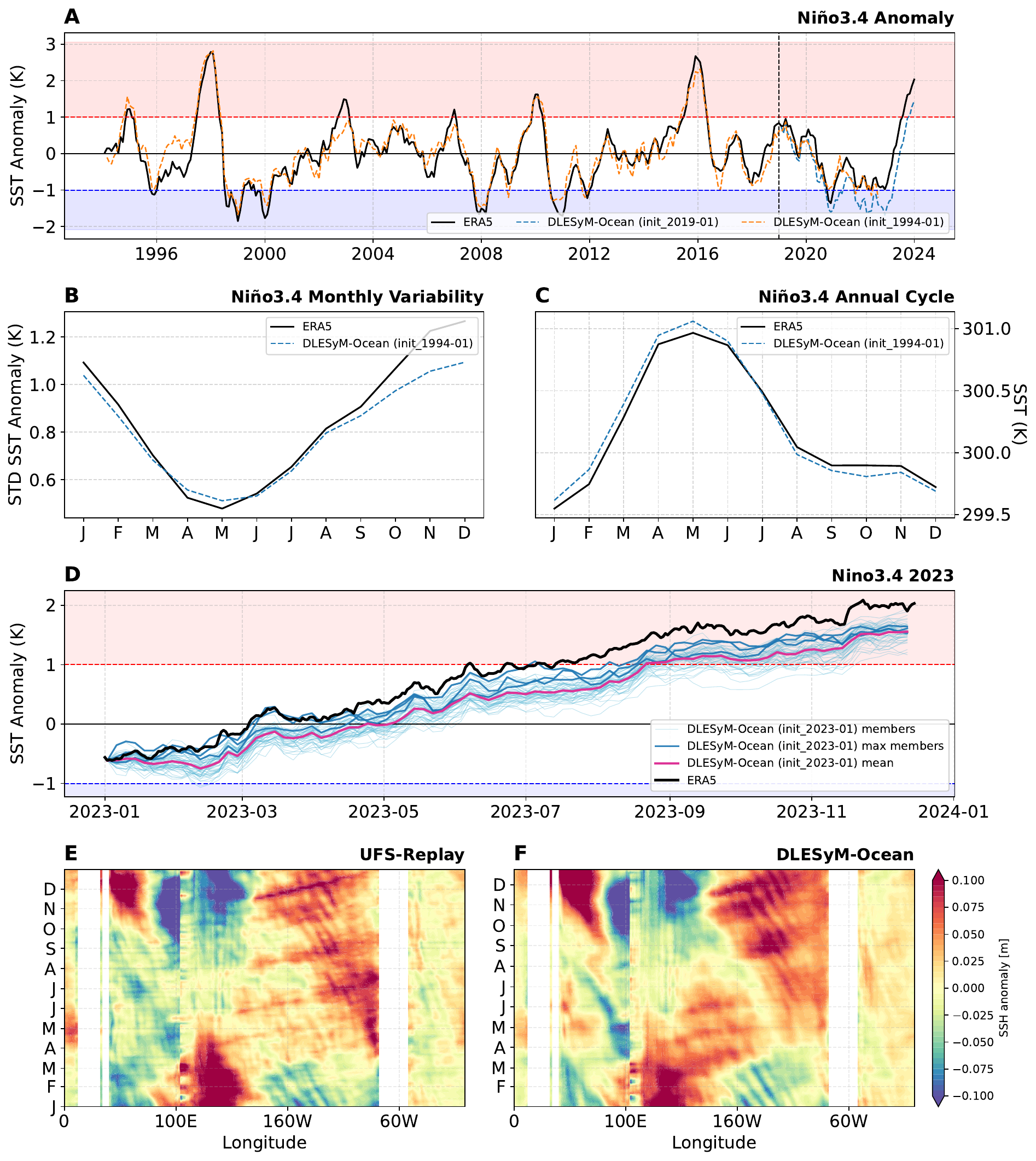}\\
 \caption{\textbf{El Niño:} (A) Time series of the sea surface temperature (SST) anomaly in the Niño-3.4 region (5$\degree$S–5$\degree$N, 170$\degree$W–120$\degree$W) from 1994 through 2023 for ERA5 (solid black line), DL{\it ESy}M-Ocean mean of ensemble simulations initialized on January 2, 1994 and forced with concurrent ERA5 reanalysis (dashed orange line; $N=16$), and DL{\it ESy}M-Ocean ensemble mean initialized in January 2019 (dashed blue line; $N=50$). Vertical dashed line separates the training period (1994–2018) from the validation and testing periods (2019–2023). Blue and red shading denote the thresholds for La Niña and El Niño events, respectively. (B) Monthly variance of the Niño-3.4 SST anomaly for ERA5 (solid black) and the ensemble mean of 1994-initialized DL{\it ESy}M-Ocean simulations (dashed blue). (C) Mean annual cycle of the raw Niño-3.4 SST for the same model configurations as in (B). (D) Time series of SST anomaly in the Niño-3.4 region 2023-01-1 through 2023-12-31 for ERA5 (solid black line), DL{\it ESy}M-Ocean ensemble mean (solid magenta line) and ensemble members (thin blue lines; $N=50$) initialized on January 1, 2023. Thick solid blue lines show the three DL{\it ESy}M-Ocean ensemble members with the lowest RMSE difference in the Niño-3.4 from ERA5 between 2023-06-01 and 2023-12-31. (E-F) Hovmöller (time-longitude) diagrams of SSH anomalies from January through December 2023 for (E) ERA5 and (F) DL{\it ESy}M-Ocean ensemble mean initialized in January 2023. All DL{\it ESy}M-Ocean simulations are forced with concurrent ERA5 reanalysis.}\label{fig:fig5}
\end{figure}

As shown in Figure~\ref{fig:fig5}A, the ensemble mean reproduces ENSO variability consistent with reanalysis, notably capturing the El Niño events in 1997/1998 and 2015/2016.  The ensemble mean for a 50-member ensemble initialized on January 1, 2019, at the beginning of the validation set,  also captures the ENSO variability, including the development of an El Niño in summer 2023. Returning to the behavior of the ensemble mean in the extended 29-year simulations, DL{\it ESy}M-Ocean generates monthly variance in SST anomalies consistent with ERA5 (Figure \ref{fig:fig5}B), although the largest systematic bias occurs during the boreal winter when monthly variability is at a maximum and SSTs are coldest. The DL{\it ESy}M-Ocean ensemble mean also captures the Niño 3.4 annual cycle quite well exhibiting very minimal drift over the 29-year driven simulation (Figure \ref{fig:fig5}C).


In 2023, ENSO transitioned from a La Niña to a near record El Niño, which helped drive the surge in global temperature \cite{raghuraman20242023}. To more closely examine DL{\it ESy}M-Ocean's treatment of this transition, which lies entirely within the test set, a second 50-member ensemble simulation was initialized on January 1, 2023.  As demonstrated by the Ni\~no3.4 anomalies plotted in Figure \ref{fig:fig5}D, DL{\it ESy}M-Ocean reproduces the rapid 2023 transition (Figure \ref{fig:fig5}D). However, no ensemble member captures the extreme amplitude of the observed ENSO anomaly by year-end, indicating a potential systematic underestimation, which likely helps explain the negative bias in global SST anomalies in 2023 (Figure \ref{fig:fig3}A). The ensemble mean of the 2023 DL{\it ESy}M-Ocean simulations also accurately captures SSH and SST perturbations in the 5\degree S to 5\degree N equatorial band and illustrated by Hovmöller diagrams (Figure \ref{fig:fig5}E-F, Figure~\ref{fig:figs9}). Wave signatures visible in the SSH field show a close match between the UFS-replay reanalysis and the DL{\it ESy}M-Ocean simulation. Eastward propagating Kelvin waves are apparent along with transequatorial tropical instability waves. 

Transequatorial tropical instability waves, which develop along the margins of the cold tongue in the region between about 100\degree W and 170\degree W, have wavelengths of roughly 1000 km and westward phase speeds of about 0.5 ms$^{-1}$\citep{Chelton_Wentz_2000}.  DL{\it ESy}M-Ocean accurately captures the SSH perturbations (Figure \ref{fig:fig5}E-F) and the thermal structure of these waves and other perturbations to a depth of 200 m (Figure~\ref{fig:figs9}).  Interestingly, precise phase relations are maintained between the reanalysis and the DL{\it ESy}M-Ocean simulation throughout the year long autoregressive rollout, suggesting that the model recognizes enough correlation between the specified atmospheric fields and the upper-ocean perturbations to keep them in phase. Although the model learns this correlation, it is not likely causal. Numerical modeling suggests the physical coupling is actually driven by the ocean, which stimulates a response in the atmospheric boundary-layer to the SST anomalies \citep{Small_Xie_Wang_2003}.

\begin{figure}
\centering
\noindent\includegraphics[width=\textwidth,angle=0]{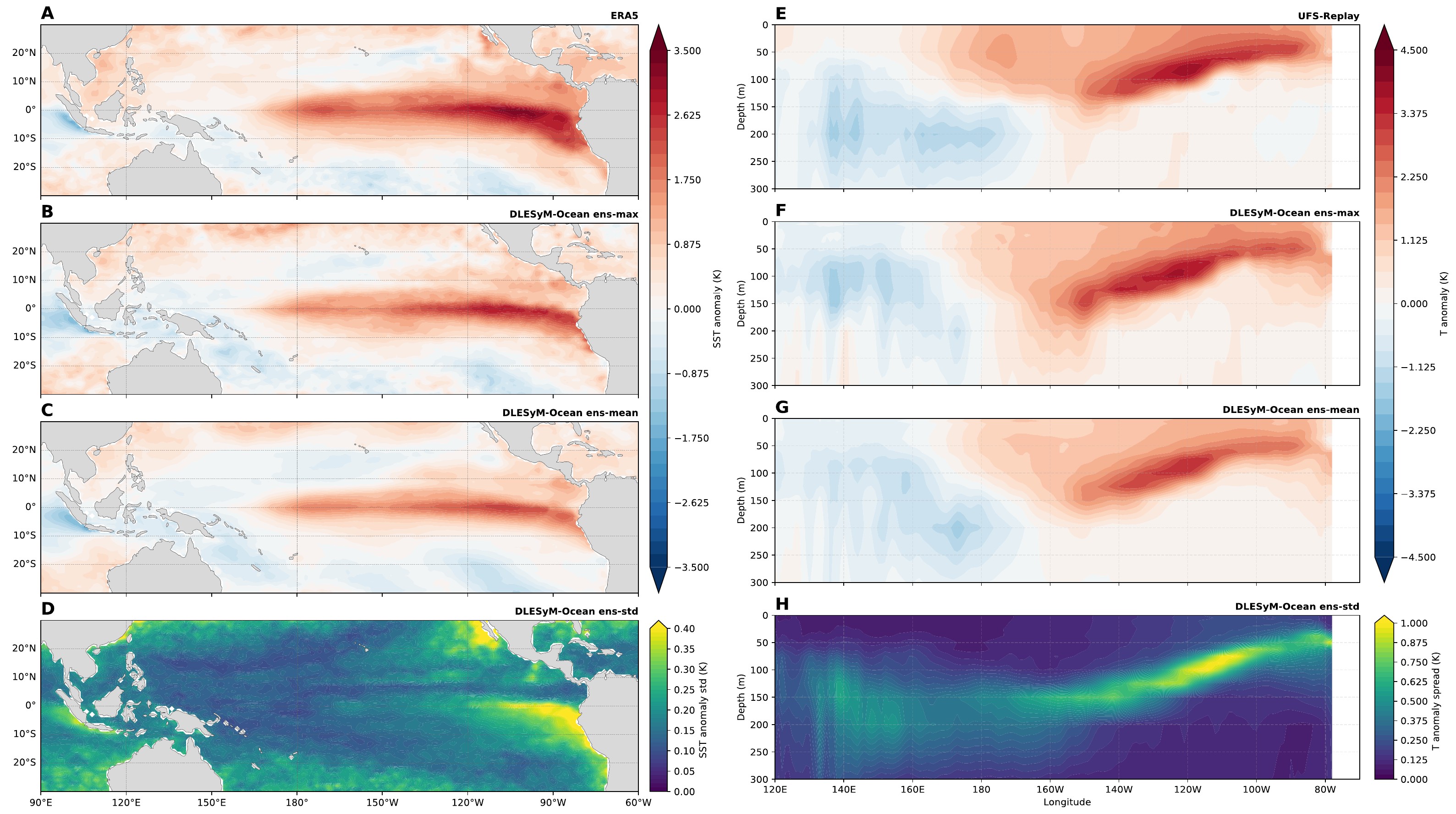}\\
 \caption{\textbf{El Niño 2023:} (A–C) Mean SST anomalies for September–December 2023 from (A) ERA5, (B) the maximum of a 50-member DL{\it ESy}M-Ocean ensemble (the member with the largest positive area-averaged Niño-3.4 anomaly), and (C) the ensemble mean. (D) Standard deviation of SST anomalies across the ensemble. (E-H) As in A-D but for subsurface temperature anomalies. All DL{\it ESy}M-Ocean simulations were initialized on January 1, 2023.}\label{fig:fig6}
\end{figure}

Figure \ref{fig:fig6} compares September–December 2023 averaged SST (A–D) and subsurface temperature (E–H; averaged 5$\degree$S-5$\degree$N) anomalies for reanalysis (A, E), the DL{\it ESy}M-Ocean member with the maximum Niño-3.4 SST anomaly (B, F), the ensemble mean (C, G), and the ensemble spread (D, H). A subset of the ensemble members are shown in Figure~\ref{fig:figs10}. The model reproduces the El Niño zonal SST dipole, characterized by pronounced eastern Pacific warming and modest western cooling. Subsurface temperature anomalies also clearly capture the corresponding eastward-shoaling zonal tilt in the equatorial thermocline. However, consistent with Figure \ref{fig:fig5}D, the magnitude of the surface and subsurface anomalies generated by DL{\it ESy}M-Ocean are biased low relative to reanalysis. The pattern of ensemble spread roughly aligns with the magnitude of surface and subsurface temperature anomalies. This indicates a well-calibrated ensemble that correctly concentrates uncertainty where prediction errors are most likely. Figures ~\ref{fig:figs11} and ~\ref{fig:figs12} show depth Hovmöller diagrams of UFS-Replay and DL{\it ESy}M-Ocean subsurface temperature anomalies for the Niño 3.4 and West Pacific regions, respectively. Each ensemble member is initialized with the same ocean state and driven by the same atmospheric forcing but follows a unique subsurface trajectory. This suggests that DL{\it ESy}M-Ocean has learned to sample plausible ocean states by emulating stochastic ocean dynamics conditioned on common atmospheric forcing.

\subsection{Variability: North Pacific Ocean Heat Wave} 
\label{sec:blob}
\begin{figure}
\centering
\noindent\includegraphics[width=\textwidth,angle=0]{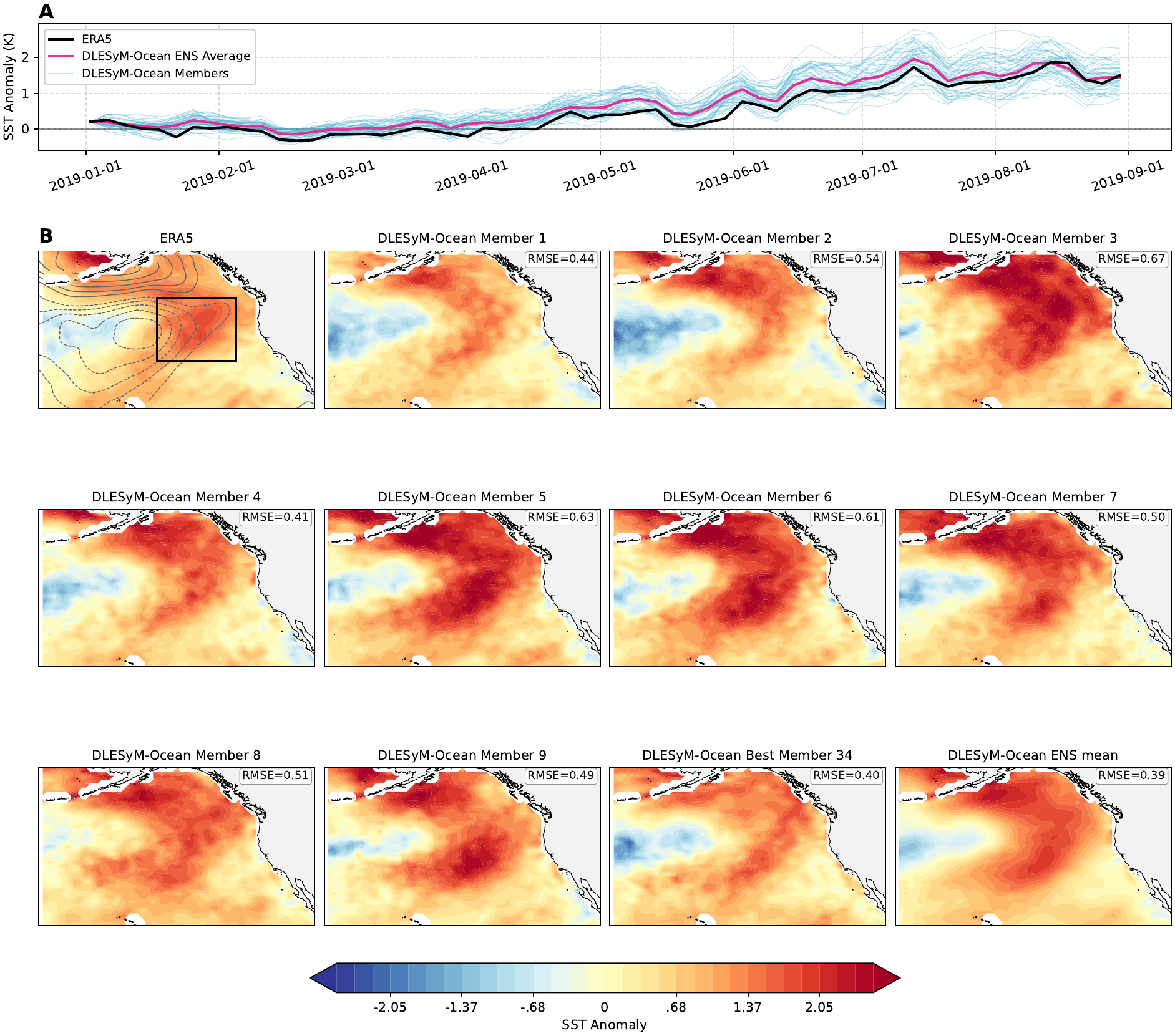}\\
 \caption{\textbf{Marine Heatwave:} (A) Time series of sea surface temperature (SST) anomaly averaged over the domain indicated by the black bounding box in (B) from January 2019 through September 2019. Data are shown for ERA5 (solid black line), the 50-member DL{\it ESy}M-Ocean ensemble mean (solid magenta line), and individual DL{\it ESy}M-Ocean ensemble members (thin solid blue lines) forced by concurrent ERA5 reanalysis. (B) Spatial distribution of ERA5 SST anomalies in the Northeast Pacific between 2019-06 and 2019-09, with the black bounding box highlighting the "Blob 2.0" marine heatwave region. All other panels are as in (B), but for 9 randomly chosen DL{\it ESy}M-Ocean ensemble members. The second to last panel shows the ensemble member with the lowest RMSE over the full domain with respect to ERA5, and the last panel show the 50-member ensemble mean. The RMSE relative to ERA5 is displayed in the upper right corner of each panel. All anomalies are computed relative to the 1994–2018 ERA5 reference climatology.}\label{fig:fig7}
\end{figure}

During the summer of 2019, anomalously warm SSTs rapidly emerged in the Northeast Pacific, an event commonly referred to as "Blob 2.0". The emergence of this marine heatwave has been attributed to a weakened North Pacific High, which induced anomalously weak surface winds, suppressed evaporative cooling and upper-ocean mixing, and increased net downward surface shortwave radiation due to reduced cloud cover \citep{amaya2020physical}. Here, we evaluate the ability of DL{\it ESy}M-Ocean to capture the magnitude and evolution of Blob 2.0 when initialized in January 2019 and integrated autoregressively through September 2019. Figure \ref{fig:fig7} presents a time series of monthly SST anomalies, spatially averaged over the region denoted by the black bounding box in Figure \ref{fig:fig7},B for ERA5 and a 50-member DL{\it ESy}M-Ocean ensemble. DL{\it ESy}M-Ocean accurately reproduces the monthly evolution of both SST (Figure \ref{fig:fig7}A) and subsurface (Figure \ref{fig:figs13}) anomalies leading up to the peak of the marine heatwave. Rows 2–4 of Figure \ref{fig:fig7} present June–September 2019 SST anomalies from ERA5 alongside a random subset of DL{\it ESy}M-Ocean ensemble members, the minimum-RMSE member, and the 50-member ensemble mean. Each member simulates a plausible marine heatwave under the prescribed atmospheric forcing, generating a spectrum of intensities that bounds the observed event. The ensemble reproduces the characteristic comma-shaped anomaly pattern, peaking off the US Pacific Northwest coast. Compared to ERA5, the forecast ensemble mean yields a pattern correlation of $r = 0.81$ and an RMSE of 0.39 K. The revealed internal variability, especially at longer lead times and below the surface, demonstrates the model's ability to autoregressively generate diverse, realistic, mostly downward-propagating trajectories of ocean heat content anomalies under common surface forcing, which may portend its eventual usefulness as a counterfactual-generator of oceanic heat extremes analogous to AI atmosphere models \citep{mahesh2025huge1,mahesh2025huge2}.

\subsection{Extremes: 2023 Austral Winter Record Low Sea Ice}
\label{sec:antarctic-low}
Following several decades of slight expansion, Antarctic sea ice extent has precipitously declined since 2016, culminating in a record low during the 2023 austral winter. During this period, Antarctic sea ice reached an unprecedented negative anomaly of 2.2 million km$^2$ relative to the historical climatology. \cite{espinosa2024understanding} reproduced this event using wind-nudging in the Community Earth System Model version 2 (CESM2), demonstrating the critical role of preceding and concurrent atmospheric conditions. Through climatological nudging, they further established that over half the magnitude of the anomaly was predictable at lead times exceeding six months. Here, we examine the ability of DL{\it ESy}M-Ocean to reproduce the 2023 austral winter record low, and assess its utility for comparable predictability and mechanistic studies.

\begin{figure}
\centering
\noindent\includegraphics[width=.9\textwidth,angle=0]{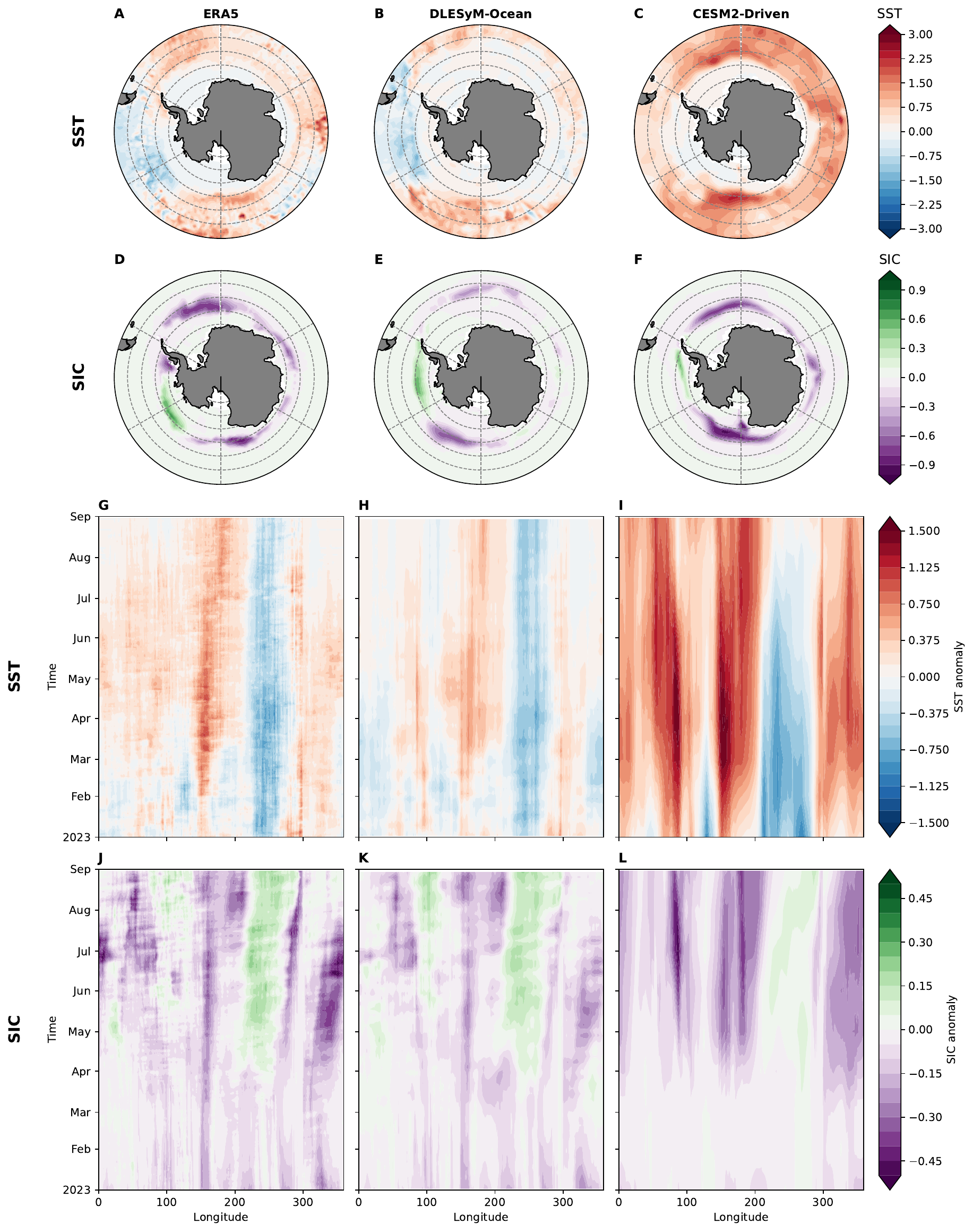}\\
 \caption{\textbf{Antarctic Sea Ice:}  (A–E) Spatial distributions of austral winter (June–August [JJA] 2023) sea ice concentration (SIC) anomalies for (A) ERA5, (B) a 50-member DL{\it ESy}M-Ocean ensemble driven by concurrent ERA5 atmospheric forcing, (C) CESM2 nudged to concurrent ERA5 winds (CESM2-Driven), (D) a 50-member DL{\it ESy}M-Ocean ensemble forecast forced by ERA5 climatology, and (E) a 22-member CESM2 ensemble forecast nudged to historical atmospheric conditions. DL{\it ESy}M-Ocean ensembles are initialized on January 1, 2023. (F–J) As in (A–E), but for SST anomalies. (K-T) Hovmoller diagrams of SST and SIC anomalies for each simulation from January 2023 to September 2023. A full description of the CESM2 simulations is detailed in \cite{espinosa2024understanding}. }\label{fig:fig8}
\end{figure}

A 50-member DL{\it ESy}M-Ocean ensemble is initialized on January 1, 2023, and integrated through September 30, 2023. The forcing by concurrent atmospheric reanalysis in the DL{\it ESy}M-Ocean simulations is comparable to CESM2 wind-nudging, with the caveat that DL{\it ESy}M-Ocean can use statistical relationships between wind and thermodynamic properties to constraint thermodynamic processes, whereas CESM2 wind-nudging explicitly separates dynamics and thermodynamics. Figure \ref{fig:fig8} compares SIC and SST anomalies from both models with the ERA5 reanalysis. The DL{\it ESy}M-Ocean ensemble-mean spatial SIC and SST patterns (Figure \ref{fig:fig8}B,E) align closely with ERA5 (Figure \ref{fig:fig8}A,D). The model captures warm SSTs and negative SIC in the Ross, Weddell, and South Pacific Oceans, alongside cold SSTs and positive SIC in the Bellingshausen and Amundsen Seas. Interestingly, the DL{\it ESy}M-Ocean anomalies are phase-shifted westward relative to ERA5 which likely emerged because the westward propagation of SIC anomalies is weaker in DL{\it ESy}M-Ocean (Figure \ref{fig:fig8}K) than in ERA5 (Figure \ref{fig:fig8}J) and CESM2-Driven (Figure \ref{fig:fig8}L). Despite this advection bias, the internal consistency between DL{\it ESy}M-Ocean SST and SIC fields remains robust. DL{\it ESy}M-Ocean anomaly patterns align more closely with ERA5 than those of CESM2-Driven, which exhibits a notably southward sea ice edge and substantially larger positive SST anomalies. The temporal evolution of anomalies in DL{\it ESy}M-Ocean also more closely agree with ERA5 than CESM2-Driven. While both DL{\it ESy}M-Ocean and CESM2-Driven exhibit persistent SST and SIC anomalies that intensify throughout 2023, CESM2-Driven displays substantial magnitude biases relative to ERA5. Importantly, CESM2-Driven is not directly initialized from ERA5 and UFS-Replay fields, and, therefore, its January 1, 2023, initial conditions differ slightly from those of DL{\it ESy}M-Ocean and ERA5 which may contribute to the substantial bias. That DL{\it ESy}M-Ocean is substantially faster and easier to initialize than ESMs makes it well-suited for this type of process-based experiment.

\section{Conclusion}
We have demonstrated that DL{\it ESy}M-Ocean closely reproduces key upper-ocean and sea-ice variables over five years of autoregressive simulation when driven by the correct atmospheric values of 1000-hPa geopotential, 10-m windspeed and OLR. DL{\it ESy}M-Ocean generates physically realistic, albeit slightly underdispersive, ensemble forecasts. It remains stable over 29-year autoregressive integrations while closely approximating climatological mean distributions of SST, SSH, and subsurface thermal and salinity structures. DL{\it ESy}M-Ocean accurately reproduces the annual cycles of Arctic and Antarctic sea-ice extent. The ensemble mean of a 29-year driven simulation well approximates the observed Ni\~no3.4 index and captures the annual cycle in both Ni\~no3.4 SST anomaly and its standard deviation. On a much shorter time scale, DL{\it ESy}M-Ocean reproduces the observed pattern of transequatorial tropical instability waves.  The model also correctly generates extreme events: the 2019 North Pacific Ocean heat wave, the 2023 El Ni\~no and the 2023 record low austral winter sea ice. We note that DL{\it ESy}M-Ocean under predicts the amplitude of the 2023 El Ni\~no, despite the skill of initialized fully-coupled dynamical forecast models in predicting this event in late 2022 \citep{blanchard2025record}. The lower skill of the model during ENSO 2023 is also noteworthy compared to its skill in simulating the 1997/98 and 2015/16 ENSO events. Given that the model is using OLR as a predictor, this suggests that the observed increase in OLR in 2023 may have been less than expected given the surface warming, or that the model is experiencing some drift. 

When driven by identical atmospheric conditions, DL{\it ESy}M-Ocean generates ensembles of diverse and realistic surface and subsurface states. This suggests that DL{\it ESy}M-Ocean is capable of sampling plausible ocean trajectories under common atmospheric conditions and may be useful for exploring counterfactual ocean states. Additional analysis is required to examine if DL{\it ESy}M-Ocean forecasts, ensemble spread, and extreme value statistics are indistinguishable from those produced by NWP ensemble forecasting systems.

The model employs a ConvNeXt U-Net architecture trained to optimize an Almost Fair Patch Energy Score (afPES) loss function and uses conditional layer normalizations to inject latent stochasticity. To overcome the limitation of CRPS, which optimizes only marginal distributions, we employ a novel application of afPES to optimize joint distributions, thereby improving spatial coherence and reducing the likelihood of spatial artifacts. An important  contribution of this work is the identification and rigorous derivation of several degeneracies in the fair CRPS and PES. One particular degeneracy in the fair CRPS has been briefly noted in the literature but not formally derived  \citep{lang2026aifs} (see Supporting Information). 

Here we have focused on the performance of DL{\it ESy}M-Ocean when coupled with ground truth atmospheric reanalysis fields of 1000-hPa geopotential height, 10-m wind speed and OLR.  By using the correct atmospheric forcing we isolated the errors in the ocean model, allowing us to assess its performance while avoiding the confounding influence of errors in the atmospheric state that would be present in forecasts with a fully coupled data-driven atmosphere-ocean model. Future work will assess the performance of such a coupled model, including coupling with either the parsimonious atmospheric module in the original DL{\it ESy}M or more complex physically constrained probabilistic variants of that model. We anticipate applying such a coupled model to both seasonal and subseasonal forecasting.



\section*{Data Availability}
All data used within this study has been obtained through publicly available data repositories. ERA5 data was downloaded from ECMWF’s Climate Data Store \citep{hersbach2023era5}. UFS-Replay data was downloaded from https://psl.noaa.gov/data/ufs\_replay/ \citep{orbe2017large, gichamo2022optimal}.


\section*{Author Contributions}
DL{\it ESy}M-Ocean Conceptualization: DRD, ZIE. Software development: ZIE, NCC, WY, PH, DP. DL{\it ESy}M-Ocean training: ZIE. DL{\it ESy}M-Ocean Execution of evaluation \& inference: ZIE. DL{\it ESy}M-Ocean Conceptualization of evaluation \& inference: ZIE, DRD, CMB, EBW, MP. Data curation: ZIE, NCC. Analysis design: ZIE, NC. Analysis execution: ZIE. Writing - main text: ZIE, DRD. Writing – materials and methods: ZIE, WY. Writing – supplement: ZIE. Figures: ZIE. 

\section*{Acknowledgments}
ZIE is supported by the U.S. Department of Energy, Office of Science, Office of Advanced Scientific Computing Research, Department of Energy Computational Science Graduate Fellowship under Award Number(s) DE-SC0023112. CMB was supported by National Science Foundation grant PLR-1936428. DRD and NCC were supported by Office of Naval Research grant N00014‐24‐12528. EBW was supported by National Science Foundation grant OPP-2213988 and OPP-2610157. This report was prepared as an account of work sponsored by an agency of the United States Government. Neither the United States Government nor any agency thereof, nor any of their employees, makes any warranty, express or implied, or assumes any legal liability or responsibility for the accuracy, completeness, or usefulness of any information, apparatus, product, or process disclosed, or represents that its use would not infringe privately owned rights. Reference herein to any specific commercial product, process, or service by trade name, trademark, manufacturer, or otherwise does not necessarily constitute or imply its endorsement, recommendation, or favoring by the United States Government or any agency thereof. The views and opinions of authors expressed herein do not necessarily state or reflect those of the United States Government or any agency thereof. This work uses ERA5 data, which contains modified Copernicus Climate Change Service information 2020. Neither the European Commission nor ECMWF is responsible for any use that may be made of the Copernicus information or data it contains. This research used resources of the National Energy Research Scientific Computing Center (NERSC), a Department of Energy User Facility using NERSC award DDR-ERCAP [0036065] and NERSC award BER-ERCAP [0037588].

\bibliographystyle{unsrtnat}
\bibliography{references}  






\newpage 

\section{Supporting Information for ``DLESyM-Ocean: A Deep Learning Probabilistic Global Model for Simulating Present-Day Upper Ocean and Sea Ice"}

\setcounter{figure}{0}          
\renewcommand{\thefigure}{S\arabic{figure}}

In this Supporting Information we include two tables and 12 figures labeled S1 to S12, referenced in the main text.

\subsection{Fair Patch Energy Score Degeneracies}
The discrete empirical formulations of both the fair Patch Energy Score (fPES) and the fair CRPS (fCRPS) suffer from critical structural degeneracies. For clarity, we illustrate these degeneracies using the scalar case ($P=1$, which reduces the fPES exactly to the fCRPS) with a two-member ensemble ($M=2$). Note that these pathologies persist in higher spatial dimensions ($P>1$) and larger ensembles ($M>2$). Recall that the discrete fCRPS is defined as:

$$fCRPS(\{x_i\}_{i=1}^M,y) = \frac{1}{M} \sum_{i=1}^M |x_i - y| - \frac{1}{2M(M-1)}\sum_{i=1}^M\sum_{j=1}^M|x_i - x_j|$$

For an ensemble of size $M=2$, expanding the summations simplifies the score to:

$$fCRPS(x_1, x_2, y) = \frac{1}{2} \Big( |x_1 - y| + |x_2 - y| - |x_1 - x_2| \Big)$$

\subsubsection*{Degeneracy 1: Bracketing Members $M=2$} Consider a scenario where the ensemble strictly brackets the target, such that $x_1 \geq y \geq x_2$. Under this condition, we can resolve the absolute values algebraically: $|x_1 - y| = x_1 - y$, $|x_2 - y| = y - x_2$, and $|x_1 - x_2| = x_1 - x_2$. Substituting these into the simplified equation yields:

$$fCRPS_{M=2} = \frac{1}{2} \Big( (x_1 - y) + (y - x_2) - (x_1 - x_2) \Big) = 0$$

Consequently, any two-member ensembles that perfectly brackets the target will evaluate to a zero score, regardless of the absolute magnitude of the errors ($x_i - y$). In a two dimensional loss landscape (i.e. $x_1, x_2$ space) the gradients in the 2nd and 4th quadrants are identically 0 everywhere. 

\subsubsection*{Degeneracy 2: Dead Gradients for Extremes $M=2$} Consider the case where the ensemble misses the target on the same side, such as $x_1 \geq x_2 \geq y$. Resolving the absolute values yields:

$$fCRPS(x_1, x_2, y) = \frac{1}{2} \Big( (x_1 - y) + (x_2 - y) - (x_1 - x_2) \Big) = x_2 - y$$

The loss reduces to the absolute error of the member nearest the target ($x_2$). The gradient with respect to the distant member ($x_1$) evaluates to zero, removing it from the optimization entirely. This dead gradient affects the first and third quadrants of the loss space. Applying the almost-fair parameterization ($afCRPS$) with $0 \leq \alpha < 1$ breaks both degeneracies (derivation omitted for brevity).

\subsubsection{Degeneracy: Unpenalized Extremes $M>2$}
Assume the ensemble is sorted such that $x_1 \leq x_2 \leq \dots \leq x_M$. If we evaluate the gradient of the fPES for patch size=1 (i.e., fCRPS) with respect to the highest member, $x_M$, the derivative of its absolute error term is $+\frac{1}{M}$ provided it overshoots the target ($x_M \geq y$):

$$\frac{\partial }{\partial x_M} \left( \frac{1}{M} (x_M - y) \right) = \frac{1}{M}$$

For the ensemble spread term, because $x_M$ is strictly greater than or equal to all other members, the absolute values resolve linearly to $(x_M - x_j)$. Its derivative is:

$$\frac{\partial}{\partial x_M} \left( \frac{1}{M(M-1)} \sum_{j=1}^{M-1} (x_M - x_j) \right) = \frac{1}{M(M-1)} \times (M-1) = \frac{1}{M}$$

Subtracting the spread gradient from the error gradient yields a total gradient identically equal to zero ($\frac{1}{M} - \frac{1}{M} = 0$). By mathematical symmetry, the total gradient for the lowest member ($x_1$) is also identically zero provided it undershoots the target ($x_1 \leq y$).This shared cancellation dictates that the fair estimator fails to penalize any extreme member extending away from the observation. Depending on the ensemble's position relative to the target, this single degeneracy manifests in three ways:

\begin{itemize}
    \item Positive Bias ($y \leq x_1 \leq \dots \leq x_M$): The upper extreme ($x_M$) experiences a zero gradient and detaches from the optimization.
    \item Negative Bias ($x_1 \leq \dots \leq x_M \leq y$): The lower extreme ($x_1$) experiences a zero gradient and detaches.
    \item Bracketing ($x_1 \leq y \leq x_M$): Both extremes experience a zero gradient simultaneously, leaving the ensemble boundaries completely unconstrained.
\end{itemize}

Unlike the $M=2$ bracketing case where this cancellation reduces the total loss identically to zero ($fCRPS_{M=2} = 0$), for $M>2$ the total fPES remains strictly positive due to the contributions of the intermediate members. However, the gradients with respect to the boundaries remain identically zero, permitting the outliers to diverge without increasing the total loss.

\begin{table}[H]
	\caption{ DL{\it ESy}M-Ocean Channels, their source dataset, weighting in the training loss, and their function as prognostic or constant fields or coupled forcing from the atmosphere. TOA is "top of atmosphere".}
	\centering
	\begin{tabular}{lllll}
		\toprule
		\cmidrule(r){1-2}
		Name     & Abbreviation     & Source & Loss Weight & Type \\ 
		\midrule
		sea ice concentration & sic & ERA5 & 2.000 & Prognostic \\
		sea surface temperature & sst & ERA5 & 0.400 & Prognostic \\
		sea ice thickness & sit & UFS-Replay & 5.882 & Prognostic \\
		sea surface height & ssh & UFS-Replay & 0.333 & Prognostic \\
		salinity 0m & s0m & UFS-Replay & 0.435 & Prognostic \\
		salinity 10m & s10m & UFS-Replay & 0.476 & Prognostic \\
		salinity 50m & s50m & UFS-Replay & 0.455 & Prognostic \\
		salinity 100m & s100m & UFS-Replay & 0.833 & Prognostic \\
		salinity 200m & s200m & UFS-Replay & 0.667 & Prognostic \\
		temperature 0m & t0m & UFS-Replay & 0.556 & Prognostic \\
		temperature 5m & t5m & UFS-Replay & 0.500 & Prognostic \\
		temperature 10m & t10m & UFS-Replay & 0.641 & Prognostic \\
		temperature 25m & t25m & UFS-Replay & 0.625 & Prognostic \\
		temperature 37.5m & t37.5m & UFS-Replay & 0.556 & Prognostic \\
		temperature 50m & t50m & UFS-Replay & 0.500 & Prognostic \\
		temperature 62.5m & t62.5m & UFS-Replay & 0.476 & Prognostic \\
		temperature 75m & t75m & UFS-Replay & 0.455 & Prognostic \\
		temperature 87.5m & t87.5m & UFS-Replay & 0.455 & Prognostic \\
		temperature 100m & t100m & UFS-Replay & 0.400 & Prognostic \\
		temperature 125m & t125m & UFS-Replay & 0.400 & Prognostic \\
		temperature 150m & t150m & UFS-Replay & 0.385 & Prognostic \\
		temperature 200m & t200m & UFS-Replay & 0.455 & Prognostic \\
		temperature 300m & t300m & UFS-Replay & 0.500 & Prognostic \\
		  geopotential height 1000 hPa & z1000-96H & ERA5 & NA & Coupled \\
		  windspeed 10m & w10-96H & ERA5 & NA & Coupled \\
		  Outgoing TOA thermal radiation & ttr-96H & ERA5 & NA & Coupled \\
		  topography & z & ERA5 & NA & Constant \\
		  land-Sea-Mask & lsm & ERA5 & NA & Constant \\
		  TOA solar insolation & I$_{TOA}$ & ERA5 & NA & Constant \\
		\bottomrule
	\end{tabular}
	\label{tab:tableS1}
\end{table}

\begin{table}[H]
	\caption{Training Procedure: All training times are reported as wall-clock time on 32 NVIDIA 40GB A100 GPUs. One epoch constitutes a pass through the full training dataset between 1994-01-01 and 2018-12-31, totaling 9131 time frames.}
	\centering
	\begin{tabular}{llll}
		\toprule
		\cmidrule(r){1-2}
		     & \textbf{Stage 1} & \textbf{Stage 2} \\ 
		\midrule
		\textbf{Number of AR steps} & 3 & 4-10 \\
		\textbf{Alpha} & 0.95 & 0.95 \\
		  \textbf{Spatial resolution} & HPX64 & HPX64 \\
		  \textbf{LR decay schedule} & Cosine & Constant \\
		  \textbf{Optimizer} & AdamW & AdamW \\
		  \textbf{Global Batch Size} & 160 & 160-32 \\
		  \textbf{EMA} & disabled & enabled \\
		  \textbf{Peak LR} & 1e-3 & 1e-6 \\
		  \textbf{Minimum LR} & 1e-6 & 1e-6 \\
		  \textbf{Linear warm-up epochs} & 5 & 1 \\
		  \textbf{Total train epochs} & 1500 & 100 for 4-10 AR \\
		  \textbf{Training Time} & 68 hours & 48 hours \\
		\bottomrule
	\end{tabular}
	\label{tab:tableS2}
\end{table}

\begin{figure}
\centering
\noindent\includegraphics[width=\textwidth,angle=0]{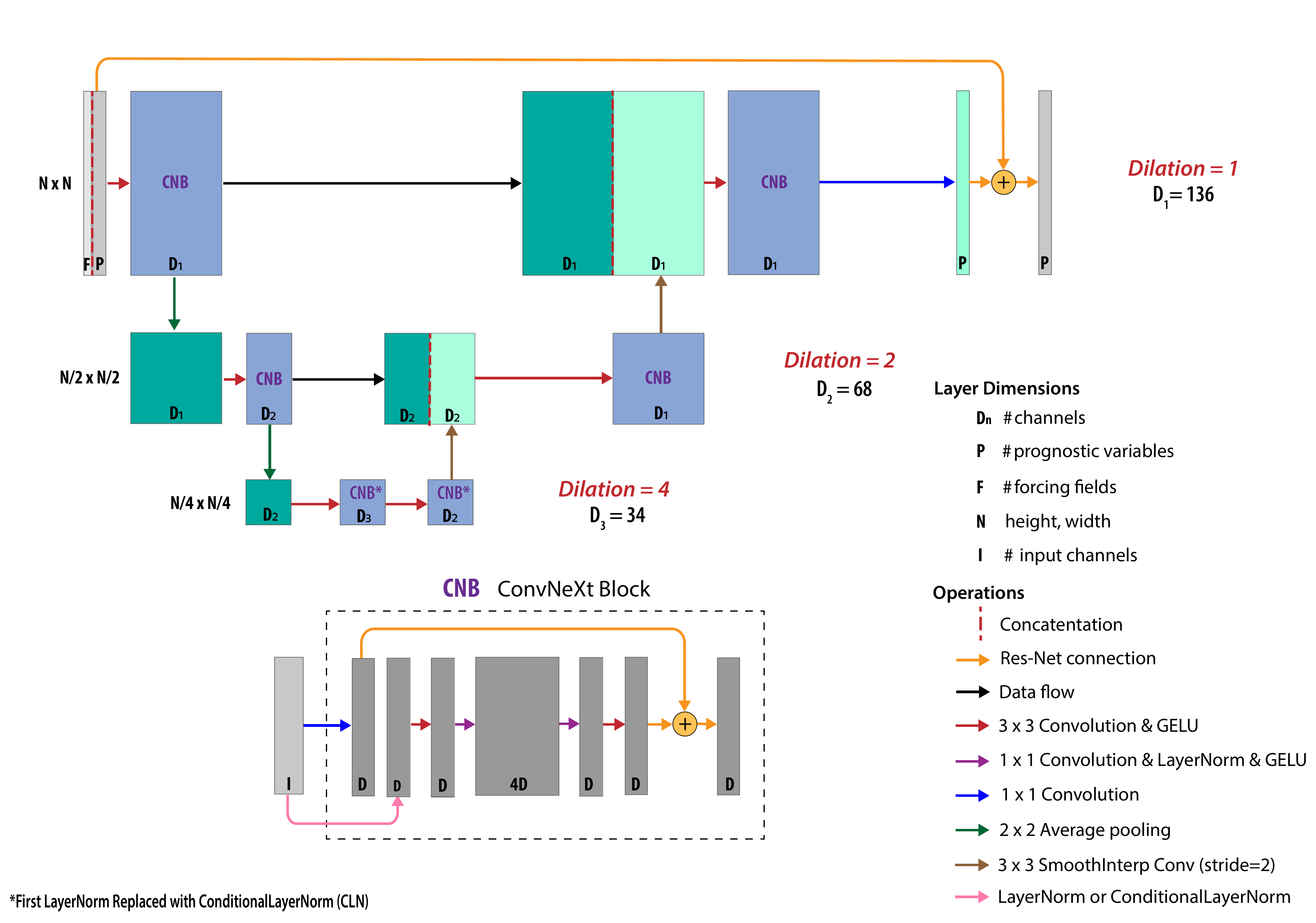}\\
 \caption{Schematic representation of DL{\it ESy}M-Ocean as a sequence of operations on layers (see legend). U-Net levels are labeled by their channel depth, with D$_1$ = 136 and D$_2$ = 68, D$_3$ = 34 being associated with the first convolutions in each level. Each ConvNeXt block (blue) is replaced by the layers and operations shown in the inset labeled CNB, with generic embedding depths D and I determined by the channel depth of the input and the labeled value of D$_n$. Other layers evaluated by the encoder are shown as dark green, while those evaluated by the decoder are shown as light green. CNB at the deepest level in the encoder and decoder labeled with * use a ConditionalLayernorm instead of LayerNorm for the first (pink arrow) normalization.}\label{fig:figs1}
\end{figure}

\begin{figure}
\centering
\noindent\includegraphics[width=\textwidth,angle=0]{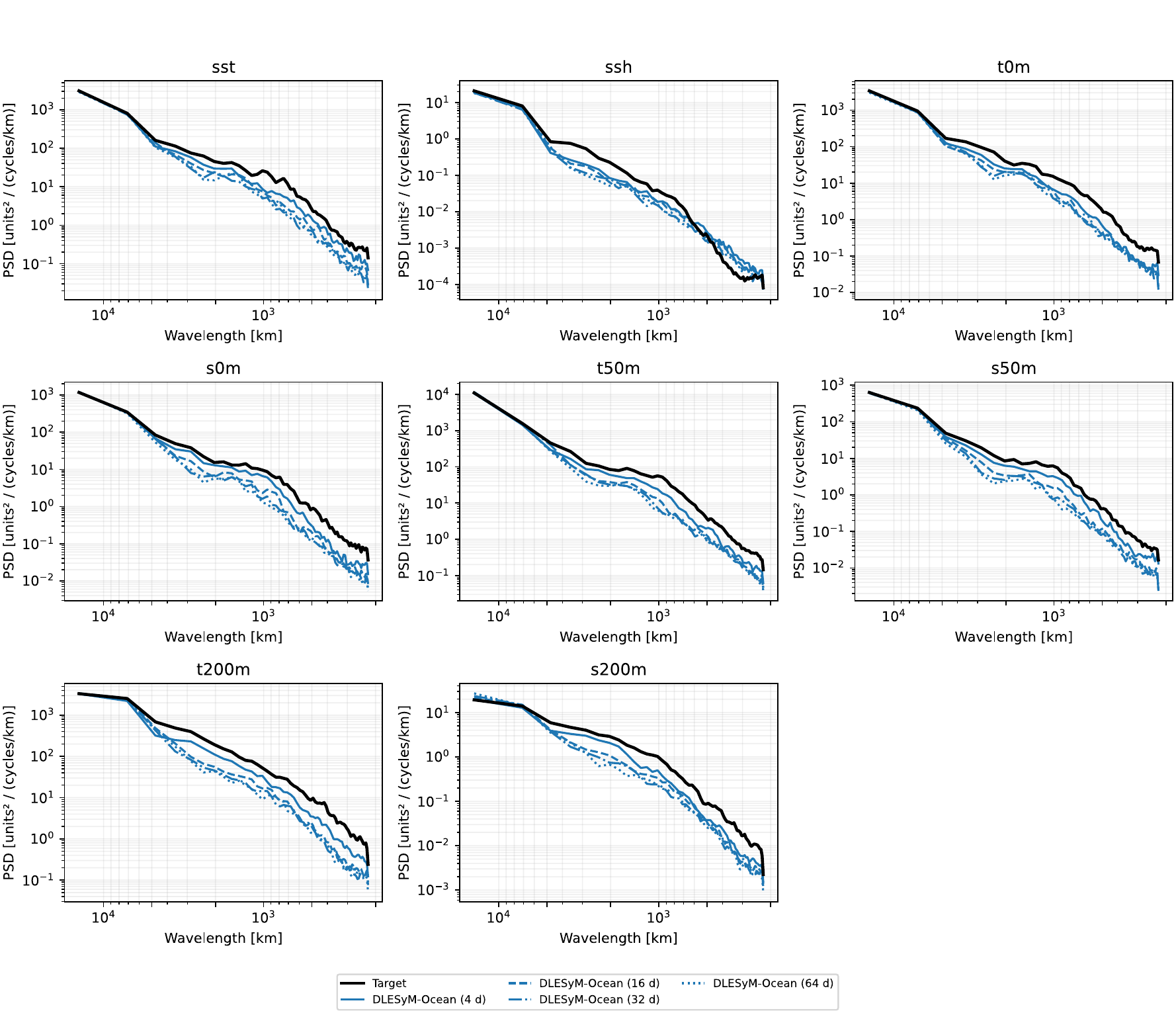}\\
 \caption{One-dimensional spatial power spectral densities (PSD) are shown for several variables and lead times using ERA5 and a 25-member, 90-day DL{\it ESy}M-Ocean forecasts initialized weekly from 2021-01 through 2023-12. The spectra are computed along a contiguous zonal transect of the equatorial (0$\degree$) Pacific. The equatorial transect is linearly detrended and a Hann window is applied prior to the discrete Fourier transform. Individual frequency spectra are computed with density scaling and averaged across all initialization times and ensemble members. The PSDs are computed at discrete wavenumbers $k$ as follows: 
  $PSD = \frac{2 \Delta x}{N \sum_{n=0}^{N-1} w_n^2} \left| \sum_{n=0}^{N-1} x_n w_n e^{-i 2 \pi k n / N} \right|^2$
 where $x_n$ is a spatial sequence of length $N$ with grid spacing $\Delta x$ and Hann window $w_n = \sin^2\left(\frac{\pi n}{N-1}\right)$.}{\label{fig:figs2}}
\end{figure}

\begin{figure}
\centering
\noindent\includegraphics[width=\textwidth,angle=0]{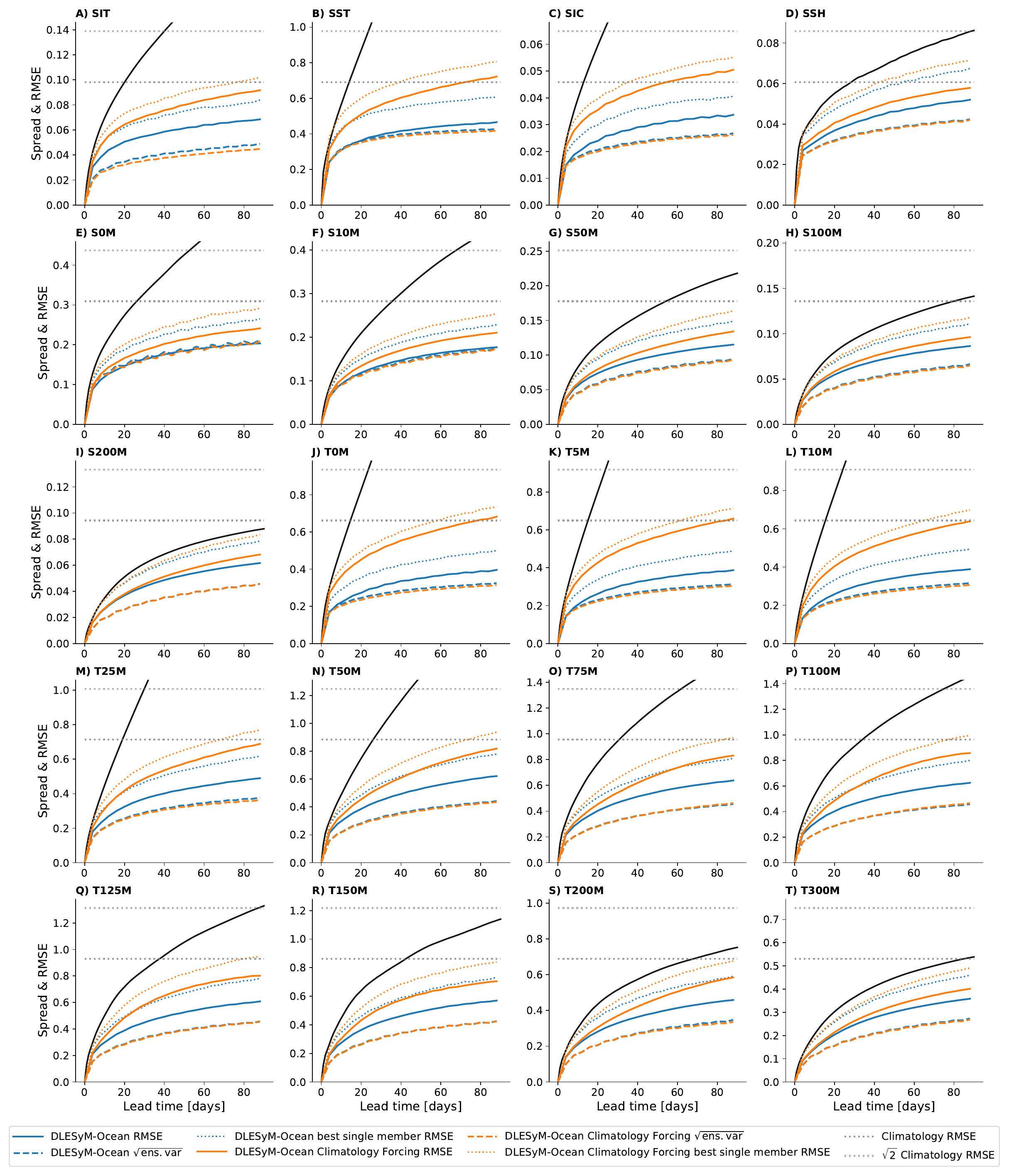}\\
 \caption{(A–T) Ensemble spread (dashed lines) and root-mean-square error (RMSE) of the ensemble mean (solid lines) for 25-member, 90-day DL{\it ESy}M-Ocean forecasts initialized weekly from 2021-01 through 2023-12. Forecasts are shown for two configurations: driven by concurrent ERA5 atmospheric forcing (blue) and driven by a daily ERA5 climatology derived from the 1994–2019 reference period (orange). Comparison benchmarks include a persistence baseline where the initial ocean state is broadcast across all future lead times (solid black line), a 25-member probabilistic climatology formulated by sampling discrete historical states from 1994–2019 (dotted gray line), and the climatological benchmark scaled by $\sqrt{2}$ (second dotted gray line).}\label{fig:figs3}
\end{figure}

\begin{figure}
\centering
\noindent\includegraphics[width=\textwidth,angle=0]{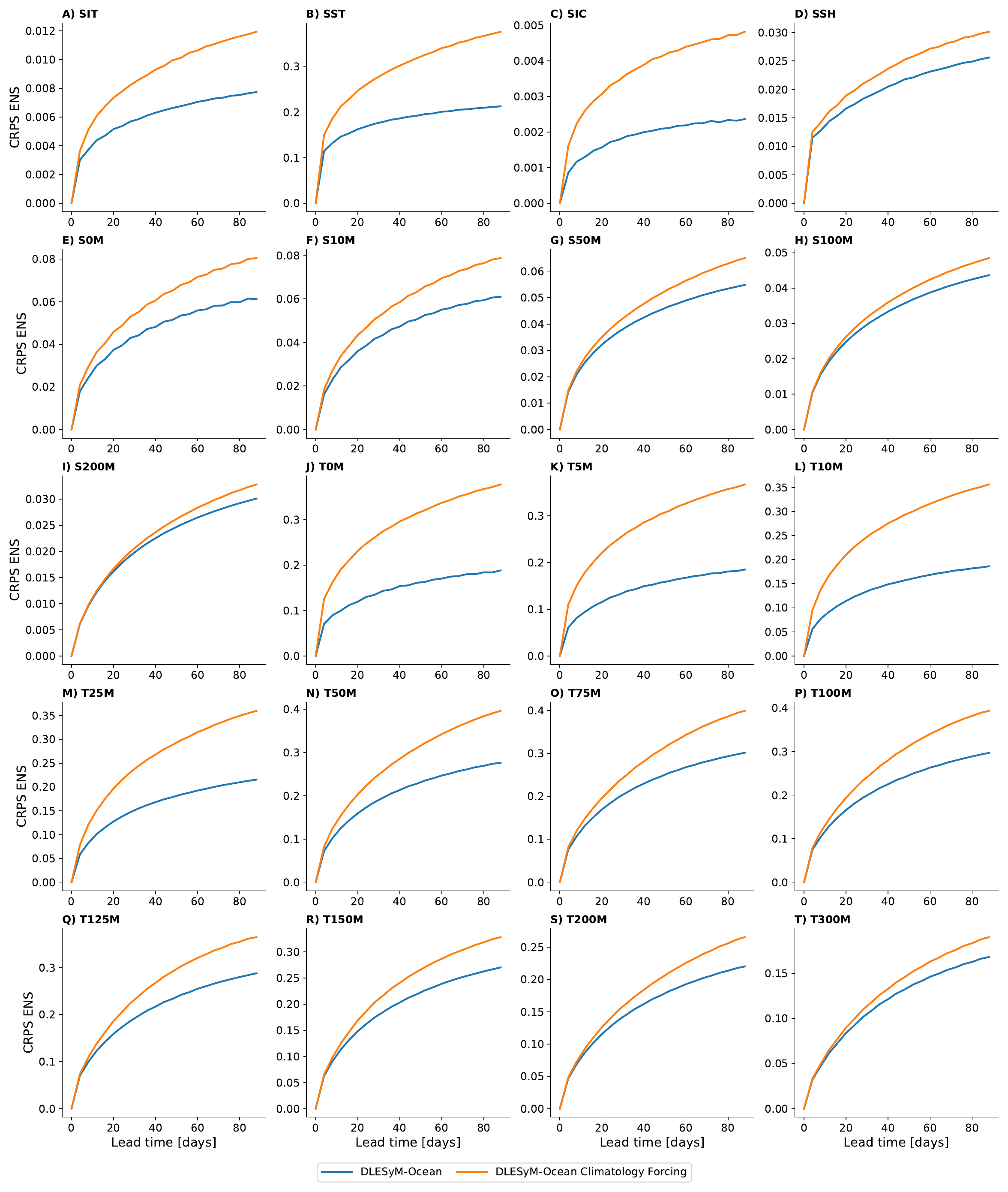}\\
 \caption{Fair CRPS for 25-member, 90-day DL{\it ESy}M-Ocean forecasts initialized weekly from 2021-01 through 2023-12 for both configurations (blue and orange lines as in Figure 1). All metrics are computed use the WeatherBenchX framework \cite{rasp2020weatherbench,raspWeatherBenchBenchmarkNext2023}. }\label{fig:figs4}
\end{figure}

\begin{figure}
\centering
\noindent\includegraphics[width=\textwidth,angle=0]{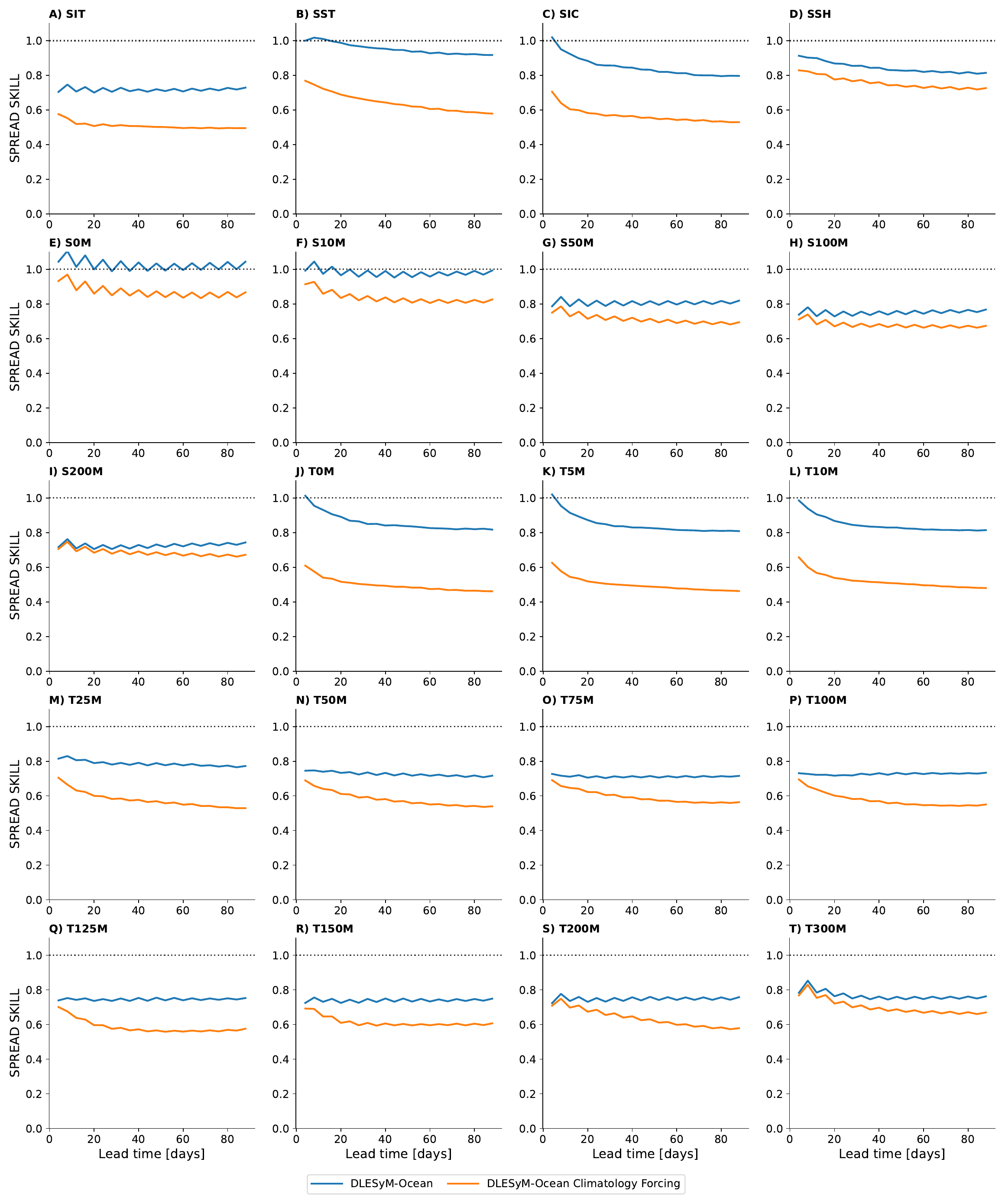}\\
 \caption{Unbiased spread-skill ratio (SSR) as a function of lead time over the 90-day forecast period for both configurations (blue and orange lines as in Figure 1). The SSR is calculated as the ratio of the ensemble spread to the RMSE of the ensemble mean (ratio of dashed and solid lines in Figure 1). A perfectly calibrated reference line is denoted at 1.0 (dotted black line). All metrics are computed use the WeatherBenchX framework \cite{rasp2020weatherbench,raspWeatherBenchBenchmarkNext2023}. To account for the finite ensemble size ($N=25$), this framework uses unbiased estimators by calculating the ensemble variance with Bessel's correction ($N-1$) and remove the positive bias in the ensemble-mean MSE by subtracting the internal ensemble variance scaled by $1/N$.}\label{fig:figs5}
\end{figure}

\begin{figure}
\centering
\noindent\includegraphics[width=\textwidth,angle=0]{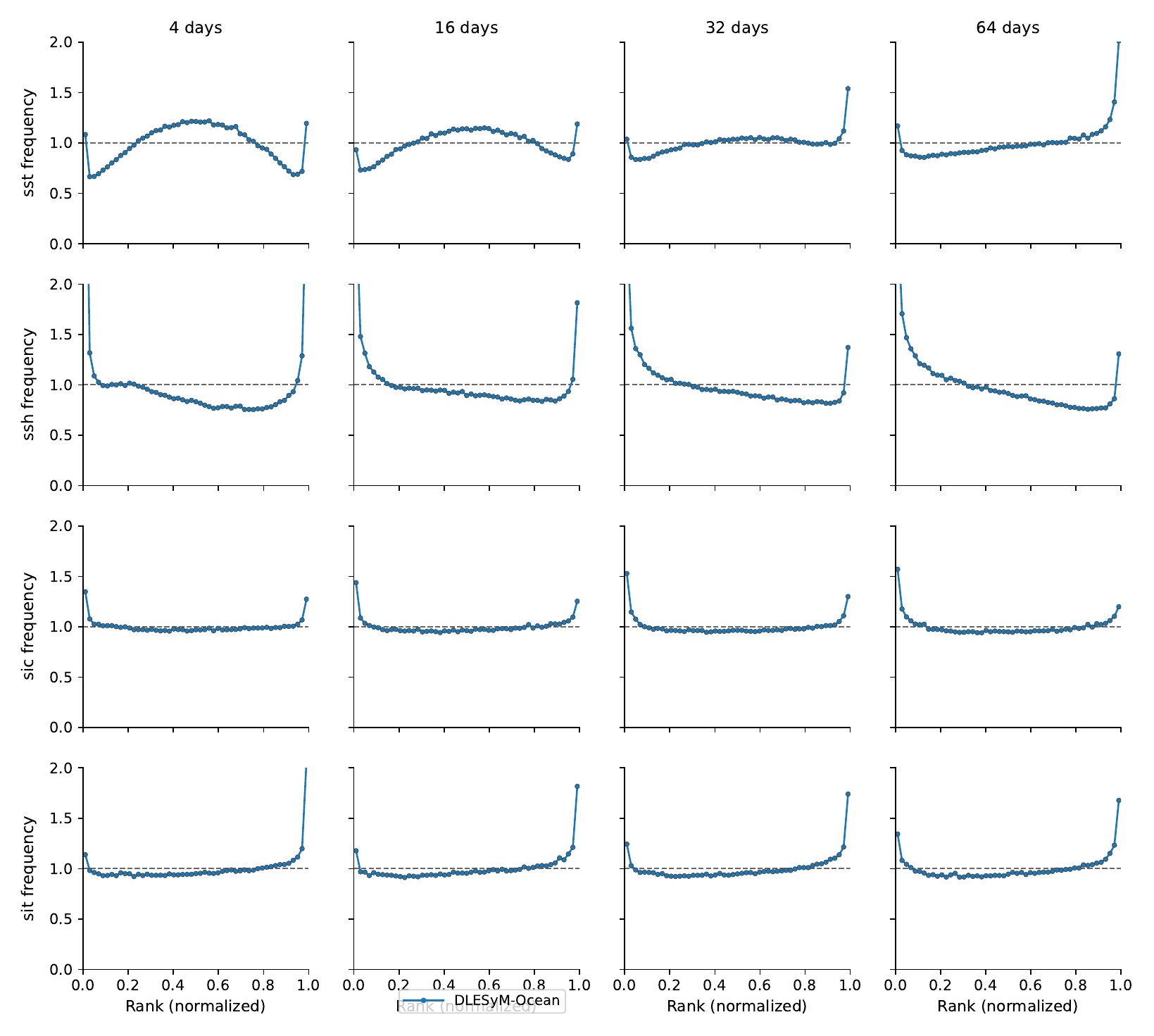}\\
 \caption{Rank histograms for 25-member, 90-day DL{\it ESy}M-Ocean forecasts initialized weekly from 2021-01 through 2023-12 for both configurations (blue and orange lines as in Figure 1). Rank histograms are constructed by tallying the frequency with which a target falls into the intervals defined by sorting the corresponding ensemble forecast members into ascending order across a sample of forecast events.  A statistically calibrated ensemble exhibits a uniform distribution, whereas U-shaped, dome-shaped, or skewed histograms indicate underdispersion, overdispersion, or unconditional systemic bias within the forecasting system, respectively.}\label{fig:figs6}
\end{figure}

\begin{figure}
\centering
\noindent\includegraphics[width=\textwidth,angle=0]{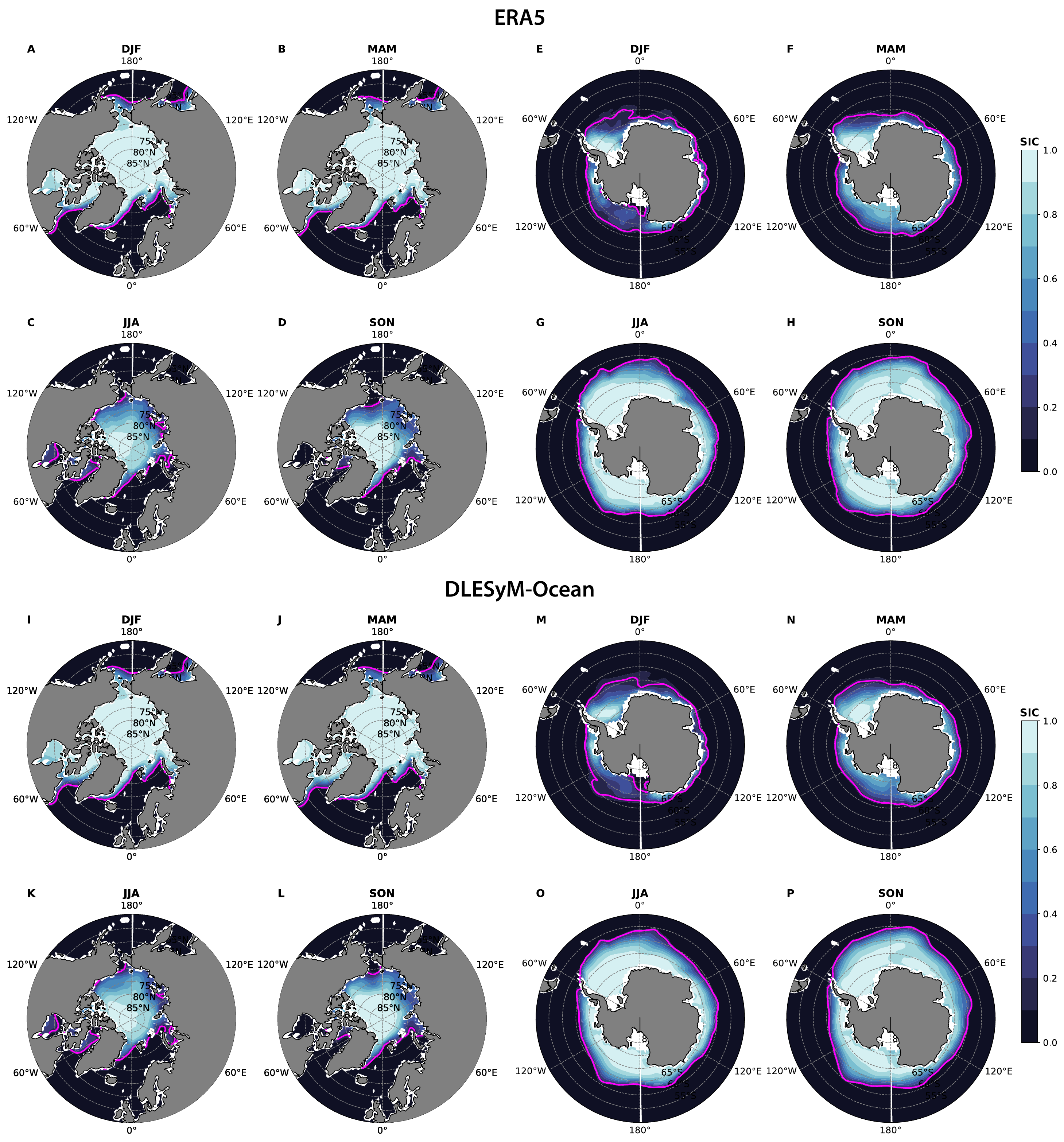}\\
 \caption{Spatial distributions of seasonal mean SIC (fractional; 0–1) for (A-D) ERA5 Arctic, (E-H) ERA5 Antarctic, (I-L) DL{\it ESy}M-Ocean Arctic, and (M-P) DL{\it ESy}M-Ocean Antarctic. Magenta contours indicate the sea ice edge (SIC = 0.15).}\label{fig:figs7}
\end{figure}

\begin{figure}
\centering
\noindent\includegraphics[width=\textwidth,angle=0]{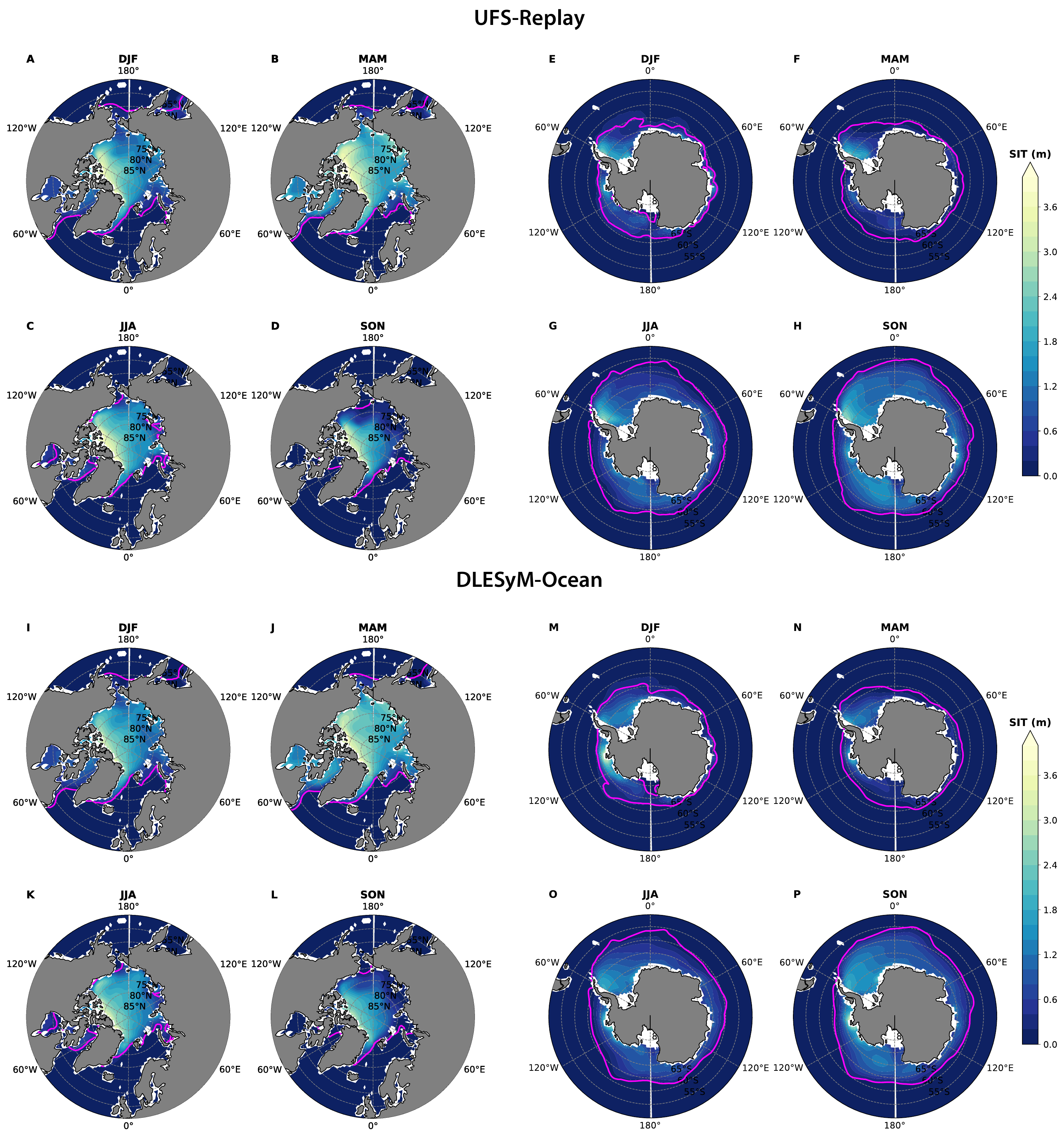}\\
 \caption{Spatial distributions of seasonal mean SIT (meters) for (A-D) ERA5 Arctic, (E-H) ERA5 Antarctic, (I-L) DL{\it ESy}M-Ocean Arctic, and (M-P) DL{\it ESy}M-Ocean Antarctic. Magenta contours indicate the sea ice edge (SIC = 0.15).}\label{fig:figs8}
\end{figure}


\begin{figure}
\centering
\noindent\includegraphics[width=.8\textwidth,angle=0]{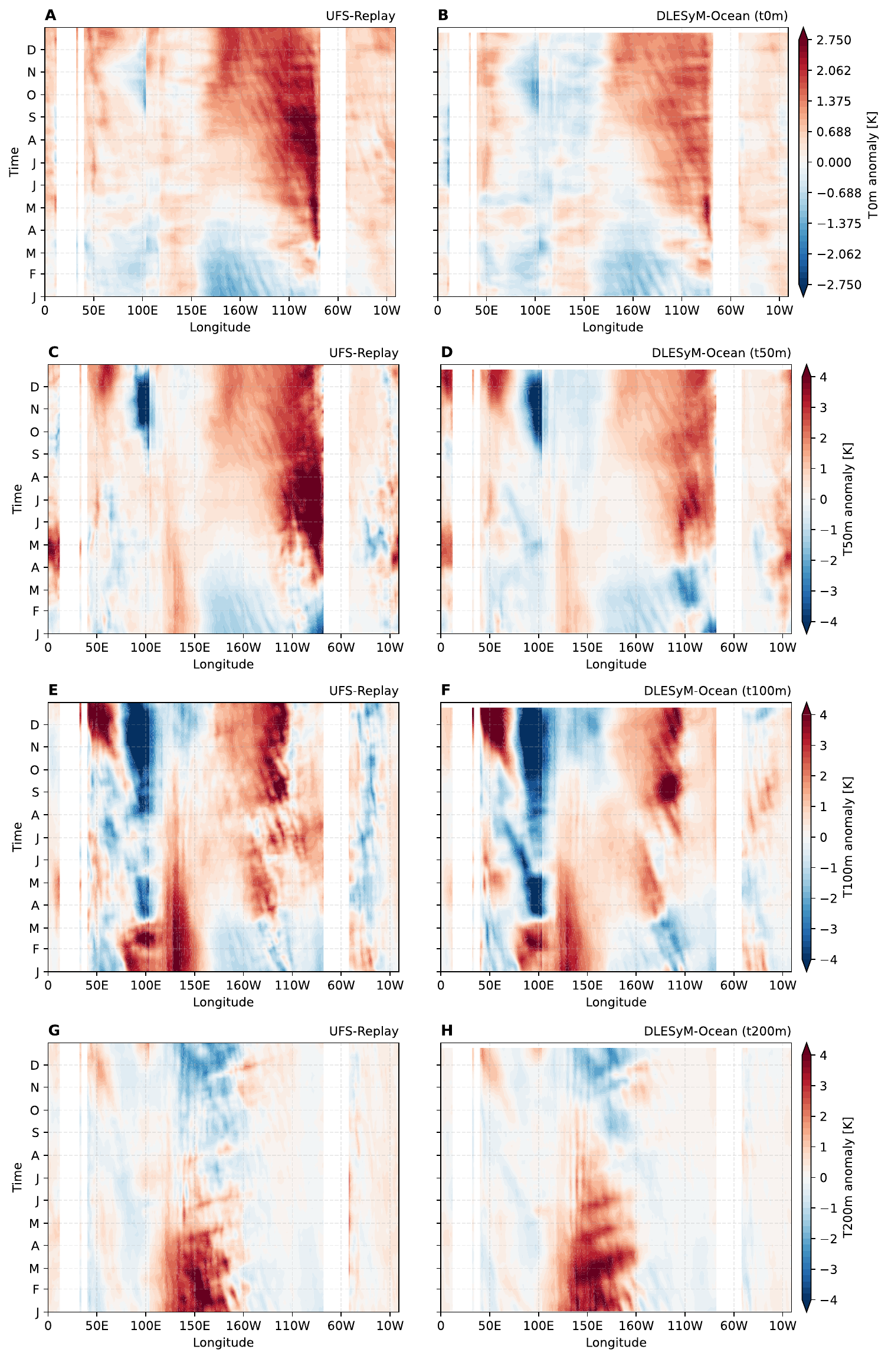}\\
 \caption{Hovmöller (time-longitude) diagrams of subsurface temperature anomalies (January–December 2023) showing the La Niña to El Niño transition. Model comparisons between UFS-Replay (A, C, E, G) and DL{\it ESy}M-Ocean ensemble mean (50-members; B, D, F, H) are presented at four depth levels: T0m (A–B), T50m (C–D), T100m (E–F), and T200m (G–H). All DL{\it ESy}M-Ocean simulations are initialized in January 2023 and forced with concurrent ERA5 reanalysis. Plotting conventions follow Figure\ref{fig:fig5}E-F.}\label{fig:figs9}
\end{figure}

\begin{figure}
\centering
\noindent\includegraphics[width=1.1\textwidth,angle=0]{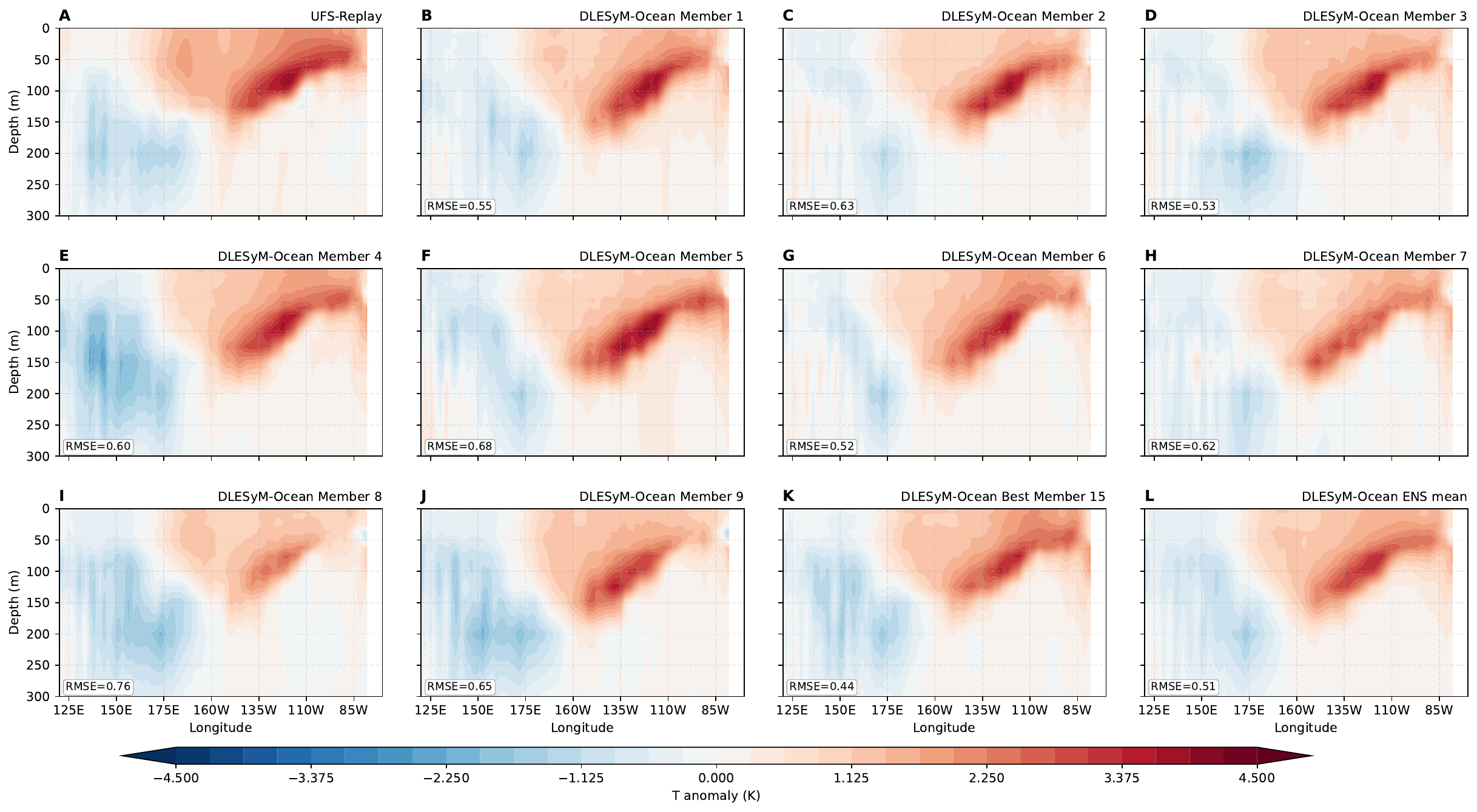}\\
 \caption{(A) Depth-Longitude of UFS-Replay subsurface temperature anomalies averaged (5$\degree$S–5$\degree$N) between September and December 2023. Panels B-J are as in (A) but for 9 randomly chosen DL{\it ESy}M-Ocean ensemble members. Panel K shows the ensemble member with the lowest RMSE over the full Hovmöller with respect to UFS-Replay, and panel P shows the 50-member ensemble mean. The RMSE relative to UFS-Replay is displayed in the lower left corner of each panel. All anomalies are computed relative to the 1994–2018 UFS-Replay reference climatology.}\label{fig:figs10}
\end{figure}

\begin{figure}
\centering
\noindent\includegraphics[width=1.1\textwidth,angle=0]{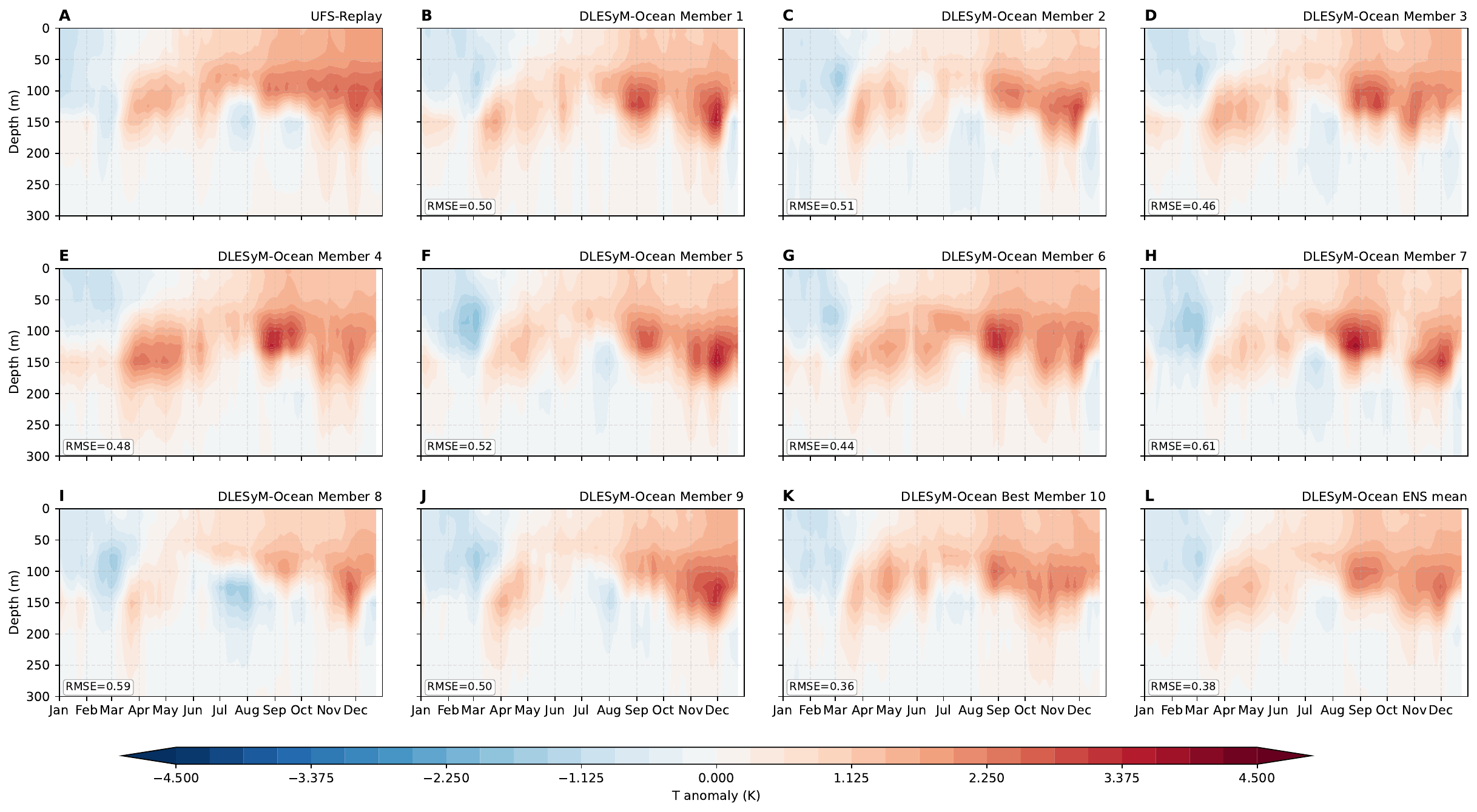}\\
 \caption{(A) Depth Hovmöller of UFS-Replay subsurface temperature anomalies area averaged in the Niño 3.4 Pacific (5$\degree$S–5$\degree$N, 190$\degree$W–240$\degree$W) between January and December 2023. Panels B-J are as in (A) but for 9 randomly chosen DL{\it ESy}M-Ocean ensemble members. Panel K shows the ensemble member with the lowest RMSE over the full Hovmöller with respect to UFS-Replay, and panel P shows the 50-member ensemble mean. The RMSE relative to UFS-Replay is displayed in the lower left corner of each panel. All anomalies are computed relative to the 1994–2018 UFS-Replay reference climatology.}\label{fig:figs11}
\end{figure}

\begin{figure}
\centering
\noindent\includegraphics[width=1.1\textwidth,angle=0]{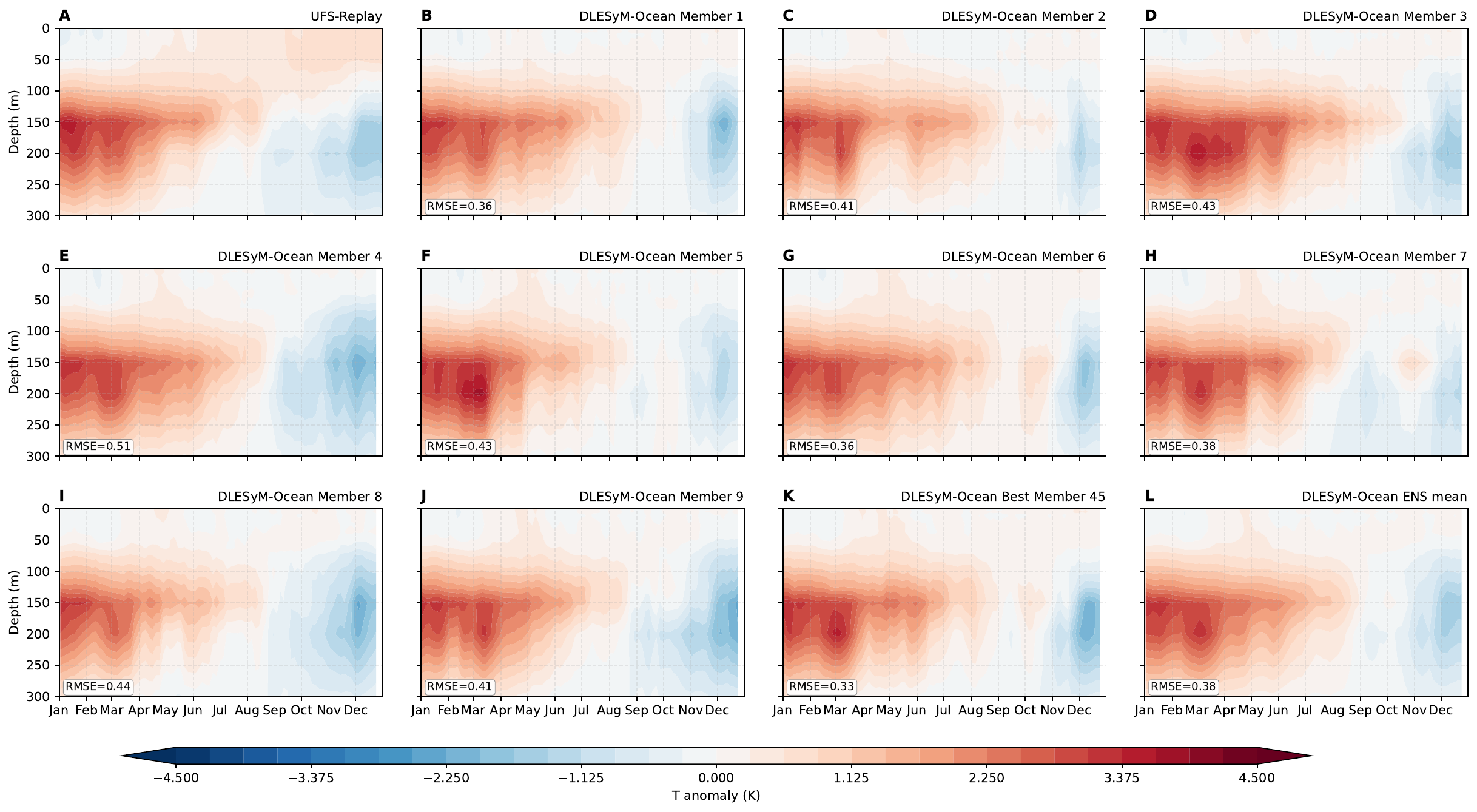}\\
 \caption{(A) Depth Hovmöller of UFS-Replay subsurface temperature anomalies area averaged in the West Pacific (5$\degree$S–5$\degree$N, 120$\degree$E–190$\degree$E) between January and December 2023. Panels B-J are as in (A) but for 9 randomly chosen DL{\it ESy}M-Ocean ensemble members. Panel K shows the ensemble member with the lowest RMSE over the full Hovmöller with respect to UFS-Replay, and panel P shows the 50-member ensemble mean. The RMSE relative to UFS-Replay is displayed in the lower left corner of each panel. All anomalies are computed relative to the 1994–2018 UFS-Replay reference climatology.}\label{fig:figs12}
\end{figure}



\begin{figure}
\centering
\noindent\includegraphics[width=\textwidth,angle=0]{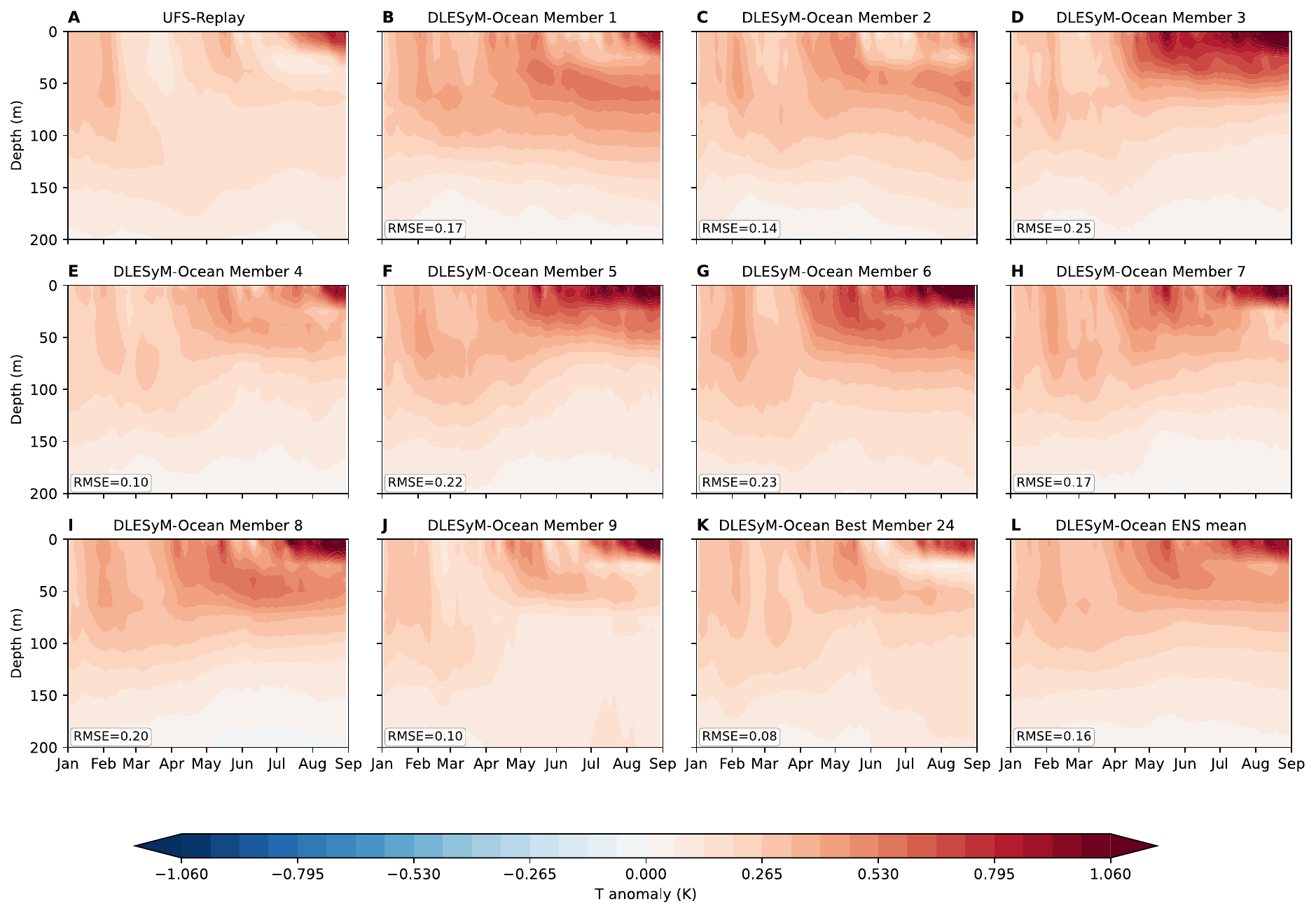}\\
\caption{(A) Depth Hovmöller of UFS-Replay subsurface temperature anomalies in the Northeast Pacific between January and September 2019. Panels B-J  are as in (A) but for 9 randomly chosen DL{\it ESy}M-Ocean ensemble members. Panel K shows the ensemble member with the lowest RMSE over the full Hovmöller with respect to UFS-Replay, and panel L shows the 50-member ensemble mean. The RMSE relative to UFS-Replay is displayed in the lower left corner of each panel. All anomalies are computed relative to the 1994–2018 UFS-Replay reference climatology. All DL{\it ESy}M-Ocean ensemble members are initialized in January and forced by the same ERA5 atmospheric conditions, such that the ensemble divergence represents plausible subsurface heat evolution due to stochastic ocean dynamics.}\label{fig:figs13}
\end{figure}


\end{document}